\documentclass{article}

\usepackage{arxiv}

\usepackage{hyperref}

\usepackage{graphicx}
\usepackage{mathptmx}

\usepackage{amsmath}
\usepackage{amsfonts}
\usepackage{amssymb}
\usepackage{amsbsy}
\usepackage{amsthm}
\usepackage{subfig}
\usepackage{tikz}
\usepackage{xspace}
\usepackage[export]{adjustbox}

\usepackage{xfrac}
\usepackage{sidecap}
\usepackage{systeme}

\usepackage{caption}
\usepackage{floatrow}   

\newcommand{\Gcmd}[1]{\texttt{#1}}
\newcommand{\stepMax}{\Gcmd{stepMax}}
\newcommand{\deltaChord}{\ensuremath{\delta_{\textrm{chord}}}}

\newcommand{\deltaIntersection}{\ensuremath{\delta_{\textrm{int}}}}

\newcommand{\dQRel}{\ensuremath{\Delta}\textrm{QRel}}
\newcommand{\dQMin}{\ensuremath{\Delta}\textrm{QMin}}
\newcommand{\deltaQ}{\ensuremath{\Delta}\textrm{Q}}

\newcommand{\ComputeStep}{\Gcmd{Compute\-Step}\xspace}
\newcommand{\OnComputeStep}{\Gcmd{OnComputeStep}}
\newcommand{\Stepper}{\Gcmd{Stepper}}
\newcommand{\AdvanceChordLimited}{\Gcmd{AdvanceChordLimited}}
\newcommand{\Interpolate}{\Gcmd{Interpolate}}
\newcommand{\reset}{\Gcmd{reset}}

\newcommand{\AccurateAdvance}{\Gcmd{AccurateAdvance}}
\newcommand{\EstimateIntersectionPoint}{\Gcmd{EstimateIntersectionPoint}}
\newcommand{\IntersectChord}{\Gcmd{IntersectChord}}

\newcommand{\GQLinkAdvanceConstrained}{\Gcmd{GQLink\_advance\_constrained}}

\newcommand{\DOPRI}{DOPRI\xspace}
\newcommand{\RK}{RK4\xspace}

\newcommand{\Speedup}{\ensuremath{\textrm{Speedup}}}
\newcommand{\SteppingSpeedup}{\ensuremath{\sigma}}
\newcommand{\CrossingSpeedup}{\ensuremath{\gamma}}

\newcommand\encircle[1]{%
  \tikz[baseline=(X.base)]
    \node (X) [draw, shape=circle, align=center, text width=1.2mm, inner sep=0] {\strut \small{#1}};}

\definecolor{lgray}{RGB}{250,250,250}

\title{Efficient Discrete-Event Based Particle Tracking Simulation for High Energy Physics}

\author{
  Lucio Santi\thanks{Corresponding author\newline\newline\copyright~2020. This manuscript version is made available under the CC-BY-NC-ND 4.0 license \url{http://creativecommons.org/licenses/by-nc-nd/4.0/}}\\
  Departamento de Computaci\'on\\
  FCEyN - UBA and ICC-CONICET\\
  Ciudad Universitaria, Pabell\'on 1\\
  C1428EGA, Buenos Aires, Argentina\\
  \texttt{lsanti@dc.uba.ar} \\
   \And
  Lucas Rossi \\
  Departamento de Computaci\'on\\
  FCEyN - UBA\\
  Ciudad Universitaria, Pabell\'on 1\\
  C1428EGA, Buenos Aires, Argentina\\
  \texttt{lrossi@dc.uba.ar}\\
  \And 
  Rodrigo Castro\\
  Departamento de Computaci\'on\\
  FCEyN - UBA and ICC-CONICET\\
  Ciudad Universitaria, Pabell\'on 1\\
  C1428EGA, Buenos Aires, Argentina\\
  \texttt{rcastro@dc.uba.ar} \\
}

\begin{document}
\maketitle
\begin{abstract}
This work presents novel discrete event-based simulation algorithms based on the Quantized State System (QSS) numerical methods. QSS provides attractive features for particle transportation processes, in particular a very efficient handling of discontinuities in the simulation of continuous systems. We focus on High Energy Physics (HEP) particle tracking applications that typically rely on discrete time-based methods, and study the advantages of adopting a discrete event-based numerical approach that resolves efficiently the crossing of geometry boundaries by a traveling particle. For this purpose we follow two complementary strategies. First, a new co-simulation technique connects the Geant4 simulation toolkit with a standalone QSS solver. Second, a new native QSS numerical stepper is embedded into Geant4. We compare both approaches against the latest Geant4 default steppers in different HEP setups, including a complex real scenario (the CMS particle detector at CERN). Our techniques achieve relevant simulation speedups in a wide range of scenarios, particularly when the intensity of discrete-event handling dominates performance in the solving of the continuous laws of particle motion.
\end{abstract}


%%%%%%%%%%% INTRO %%%%%%%%%%%%%
\section{Introduction}
\label{sec:intro}

Particle tracking simulations in the High Energy Physics (HEP) domain deal with reproducing trajectories of subatomic particles affected by physics processes within complex detector geometries, typically composed by adjacent 3D volumes of different shapes and materials. Modern HEP experiments usually rely upon the Geant4 simulation toolkit \cite{agostinelli2003geant4, G42016} to carry out these simulations. 

One of the most important applications of HEP simulations is to drive the design and optimization of particle detectors for best physics performance. In this scenario, different detector parameters are varied in the simulation phase and the actual design is chosen following a trade-off between monetary cost and detector performance. In the last decades, this simulation phase has become a strong requirement for HEP experiments to apply for funding. One of such experiments is the Compact Muon Solenoid (CMS) particle detector \cite{CMS} at the Large Hadron Collider (LHC) \cite{LHC} at CERN, for which simulation has taken approximately 85\% of the total CPU time utilized by the experiment since the start-up in 2009 through May 2016, with a total of about 40\% for the Geant4 module \cite{HEPChallenges}. Assuming a cost of 0.9 US dollar cents per CPU core hour, the annual simulation cost of the CMS experiment is in the range of 3.5 to 6.2 million US dollars. 

%In consequence, even a 1\% improvement in the Geant4 simulation module would yield from 50 to 80 thousand US dollars per year of savings to CMS.

As the needs of more precision and more varied experiments grow, it becomes a necessity to find ways of improving the efficiency of simulations, where even small percentages of improvement can yield considerable impacts in terms of savings, and therefore in the palette of simulations affordable within given budget and time constraints. In this work, we approach the problem from the perspective of the underlying numerical algorithms that solve the ordinary differential equations (ODEs) that govern the movement of particles.

Standard particle tracking algorithms in Geant4 rely exclusively on classical numerical methods, i.e. those that are based on some form of time discretization \cite{Cellier}. In particular, variations of the Runge-Kutta (RK) family of solvers \cite{butcher} are the most used ones. Yet, as we shall see later in detail, particle tracking within geometries is a peculiar problem that deals with frequent discontinuities caused by the recurrent crossing of boundaries by the particle, from one volume to the next. Classical numerical methods are not naturally prepared to deal efficiently with such situations. They must interrupt the standard integration procedure and invoke (usually costly) iterative procedures to detect the time and values of state variables at the instant of each discontinuity. %Can this situation be tackled in another way?

On the other hand, Quantized State System (QSS) is a family of methods that  discretize the state variables instead of slicing time, and solve ODEs using discrete--event (rather than discrete--time) approximations of continuous models \cite{Cellier,Kof01j}. QSS are hybrid numerical methods that combine continuous dynamics with discrete-event dynamics to approximate continuous systems. A feature of QSS that stands as very attractive in the context of HEP simulations is that the methods handle discontinuities (such as volume crossings) very efficiently \cite{kofman2004discrete} by means of a computationally cheap procedure. Discontinuities are solved using simple zero-crossing polynomials, which are treated as discrete--events for which the methods are prepared by definition.

\subsection{Methodological approach}
\label{sec:methodological_approach}

Our research is mainly driven by the question of whether discrete--event numerical integration methods (in particular, the QSS family) can be correctly and soundly used to simulate HEP experiments. As explained above, due to the nature of these experiments, their simulations face challenges that match some inherent features of discrete--event solvers --primarily, efficient handling of discontinuities. As such, we are also interested in investigating whether HEP simulations can profit from these features, and to what extent. In order to accomplish these goals, this work is structured as follows:

\begin{itemize}
    \item The starting point is to address the feasibility of QSS in the HEP domain by means of a simple tracking experiment taken as baseline. This is achieved by comparing the state-of-the-art implementation of QSS methods, QSS Solver \cite{FK14}, and Geant4, the most representative HEP simulation toolkit, in the simulation of a simple 2D-oscillating charged particle in a uniform magnetic field.
    
    \item The next step is to interface the two simulation engines so that Geant4 can rely upon QSS Solver to compute particle trajectories. This would enable not only more direct and fair comparisons but also the possibility of studying QSS in more realistic HEP setups (i.e. with complex detector geometries and physics interactions). For this purpose, we introduce a Geant4-to-QSS Solver Link (GQLink), a co-simulation technique where QSS Solver takes over the particle propagation responsibilities typically handled by native Geant4 steppers and integration drivers. This approach also aims to demonstrate that a smooth coupling between Geant4 and QSS can be achieved.
    
    \item We then move on to develop optimized QSS steppers implemented as extensions to Geant4, eliminating the interconnection layer required by GQLink and its potential performance penalties. The goal is to assess,  and eventually reduce, the overhead introduced by the co-simulation interface.
    
    \item Both QSS approaches (co-simulation and native stepper) are analyzed by studying their performance against the most common Geant4 steppers in two complementary scenarios: a 3D extension of the simple 2D oscillating particle  experiment taken as baseline, and a very complex and realistic standalone Geant4 application modeling the CMS particle detector. The main goal is to verify that QSS can indeed offer performance gains in scenarios with heavy volume crossing activity, but we also seek to provide a more general characterization of both strategies that could eventually serve as a guideline for the Geant4 community to identify other setups in which these solutions could be efficiently applied.
\end{itemize}

The rest of the paper is organized as follows: in Section~\ref{sec:background} we present an overview of relevant fundamental concepts, including the numerical solution of continuous systems by means of classical, discrete-time methods as well as discrete-event methods, an introduction to the chosen simulation toolkits and a summary of related work in the field. We continue by describing the two complementary case studies that will be used as performance benchmarks in Section \ref{sec:casestudies}. Then, Section \ref{sec:chep} reviews the initial proof-of-concept performance comparison between QSS Solver and Geant4 in a basic HEP setup that provided the foundations for the upcoming development efforts. Next, Sections \ref{sec:cosimulation} and \ref{sec:stepper} introduce our two alternative implementations of QSS methods in Geant4: GQLink, a generic co-simulation technique that connects Geant4 with QSS Solver, and a native Geant4 stepper that builds upon GQLink's core algorithms in order to boost its performance. Each Section includes detailed performance comparisons against the two most relevant Geant4 steppers. Finally, Section~\ref{sec:conclusions} provides a summary, conclusions and comments on our work in progress.

%%%%%%%%%%% BACKGROUND %%%%%%%%%%%%
\section{Background}
\label{sec:background}

In this section we present the essential concepts used along the article. We first discuss the numerical solution of continuous systems focusing on the fundamental differences between classical, discrete-time methods and  discrete-event based methods such as the Quantized State System family. We continue with a brief summary of QSS Solver, a simulation toolkit that provides state-of-the-art implementations of the QSS numerical methods, and Geant4, the most widely used simulation toolkit in HEP experiments. Finally, we summarize relevant related work in the field. 

\subsection{Numerical solution of continuous systems}
\label{sec:odes}

A dynamic \textsl{continuous system} is made up of state variables that change continuously over time. They are typically described by differential equations, involving either ordinary (ODEs) or partial (PDEs) derivatives. Since it is generally not possible to find closed-form, analytic solutions to these systems, it is important to simulate them numerically by means of numerical integration methods. 

In what follows, we will consider a system of ODEs in the form of Equation \ref{eq:ODE}, where $\mathbf{x}(t)$ is the \textsl{state vector} and $\mathbf{u}(t)$ is the \textsl{input vector} representing independent variables for which no derivatives are present in the system.
\begin{align} 
  \dot{\mathbf{x}}(t) = f(\mathbf{x}(t), \mathbf{u}(t)) \label{eq:ODE}
\end{align}    
The component $x_i(t)$ of the state vector represents the $i$-th state trajectory as function of time, which will be continuous as long as there are no discontinuities in $f_i(\mathbf{x}, \mathbf{u})$. Such discontinuities, in case they are present, must be handled carefully and efficiently during the simulation in order to preserve the accuracy of the results.

\subsubsection{Classical discrete time methods}
\label{sec:rk}

Traditional discrete-time methods solve continuous systems such as the one in the form of Equation \ref{eq:ODE} making use of \textsl{time slicing}: given the current and past values of state variables and their derivatives, the solver estimates the next state value one ``time step'' $\Delta t$ in the future (i.e., at $t_{k+1}=t_{k}+\Delta t$), where $\Delta t$ applies to all state variables and represents the means to control the approximation accuracy. This is achieved by evaluating the Taylor expansion of $x_i$ around $t_k$:
\begin{align} 
  x_i(t_{k+1}) = x_i(t_k) + \dot{x_i}(t_k) \cdot \Delta t + \ddot{x_i}(t_k) \cdot \frac{\Delta t^2}{2!} + \dots  \label{eq:taylor}
\end{align}    
Integration algorithms differ in how they compute the higher-order derivatives of the state variable and how they approximate (truncate) the infinite Taylor series (which gives rise to the \textsl{accuracy order} of the method). Runge-Kutta (RK) \cite{butcher} is a widely adopted family of numerical solvers that compute the derivatives through a series of \textsl{stages}, each one performing an evaluation of the right-hand side of Equation \ref{eq:ODE}. The fourth-order accurate Runge-Kutta (\RK) is one of the most popular of such methods. The Dormand-Prince (\DOPRI) method \cite{Dormand1980Family} is another RK-based algorithm involving six function evaluations to calculate fourth- and fifth-order accurate solutions.

In order to achieve certain accuracy constraints with minimum computational effort, these methods are usually implemented with some sort of adaptive step size control. In the context of Geant4, for example, the original \RK method is supplemented with \textsl{step doubling} \cite{NumericalRecipes1992}. With this technique, each step is taken twice: once as a full step and then, independently, as two half-steps.

When faced with discontinuous systems, these methods usually introduce ad-hoc solutions to properly locate and handle the discontinuities, which is essential to avoid integrating past a discontinuity since this may lead to incorrect results. These custom, iterative algorithms can be computationally expensive.

\subsubsection{Discrete-event methods and Quantized State System (QSS)}
\label{sec:qss}

As opposed to discrete-time integration methods, discrete-event methods discretize the state variables rather than slicing time. The most salient family of discrete-event based numerical methods is QSS, which operates by means of \textsl{state space quantization}: QSS calculates the smallest time step $h$ in the future at which the state variable differs from its current value by one ``quantum level'' $\deltaQ$, i.e., when $x_{t_k+h} = x_{t_k} \pm \deltaQ$. The system described by Equation \ref{eq:ODE} is thus approximated by the quantized system shown in Equation \ref{eq:ODEQSS}, where $\mathbf{q}(t)$ is the \textsl{quantized state vector} resulting from the quantization of the state variables $x_i(t)$. 
\begin{align} 
  \dot{\mathbf{x}}(t) = f(\mathbf{q}(t), \mathbf{u}(t)) \label{eq:ODEQSS}
\end{align}    
In the first-order QSS method (QSS1) each $q_i(t)$ follows a piecewise constant trajectory that is updated by a (hysteretic) \textsl{quantization function} when the difference between $q_i(t)$ and $x_i(t)$ reaches the \textsl{quantum} $\deltaQ_i = \textrm{max}\left(\dQRel \cdot |x_i|, \dQMin\right)$ derived from the precision demanded by the user by means of a relative and a minimum quanta, $\dQRel$ and $\dQMin$. In QSS1, $\mathbf{q}(t)$ follows piecewise constant trajectories, which implies that $\mathbf{x}(t)$ follow piecewise linear trajectories. Along the same principle, higher-order QSS methods generalize this behavior: in QSS$n$, $\mathbf{x}(t)$ follow piecewise $n$-th degree polynomial trajectories and $\mathbf{q}(t)$ follow piecewise $(n-1)$-th degree polynomial trajectories \cite{Kof01j}.

\begin{figure}[h!]
\centerline{
    \includegraphics[scale=0.7]{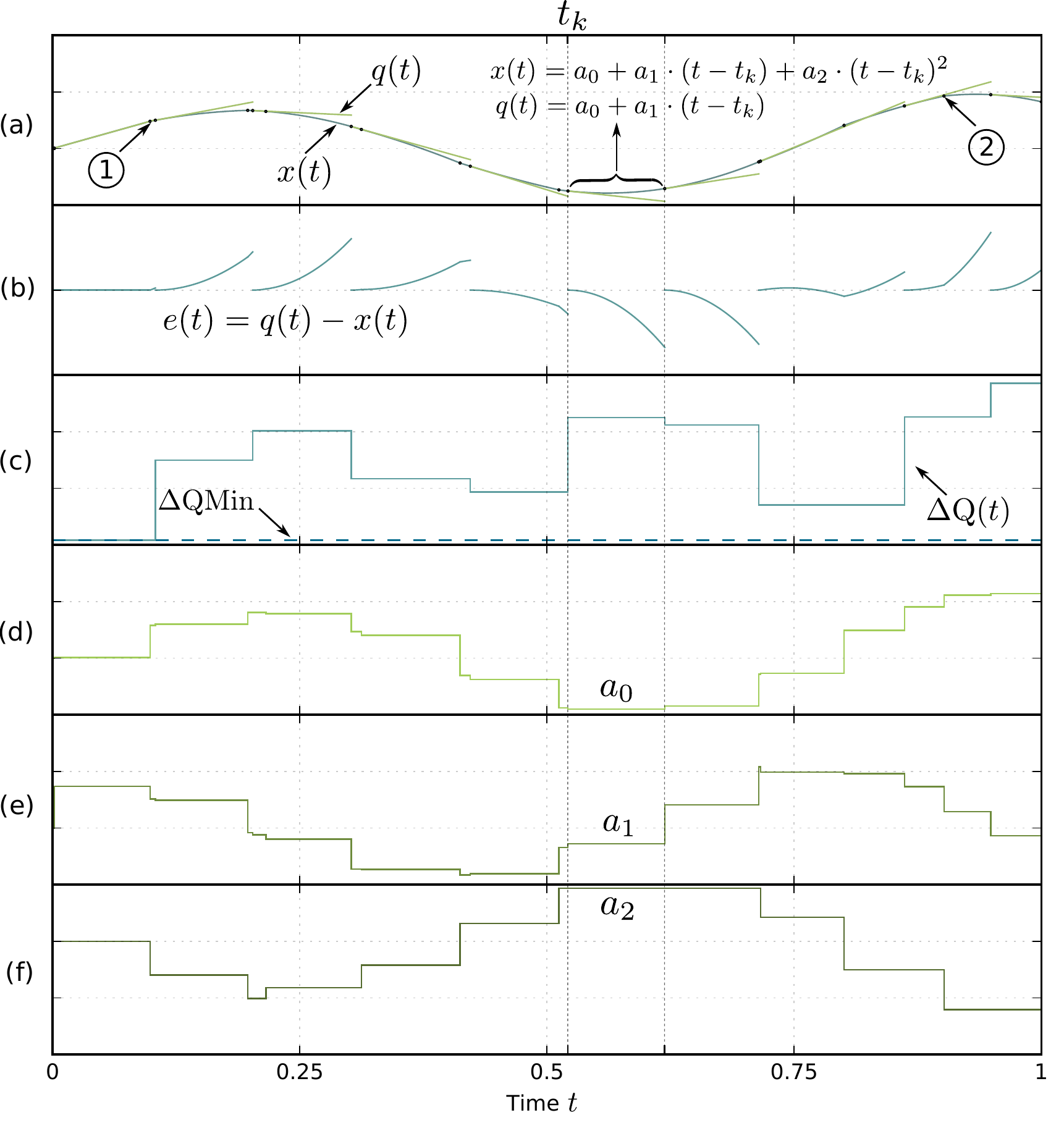}
}
    \caption{Illustrative example of a QSS2-based simulation and its main underlying QSS concepts}
    \label{fig:qssplot}
\end{figure}

These concepts can be visualized in Figure \ref{fig:qssplot}, which presents a QSS2 simulation of the position in the $\hat{x}$ axis of a charged particle in a constant  magnetic field (the model in use is the ODE system in Equation \ref{eq:lorentz}, which will be revisited in Section \ref{sec:chep_impl}). Figure \ref{fig:qssplot}a shows the solution state variable $x(t)$ and its corresponding quantized state variable $q(t)$, which follow piecewise quadratic and linear trajectories, respectively. Each dot in the curve marks the ending and the commencements of adjacent sections. Sections are limited either by reacting to an update originated from another state variable (e.g. \encircle{1}) or by reaching the quantum $\deltaQ$, i.e. the maximum deviation allowed between $q(t)$ and $x(t)$ (e.g. \encircle{2}). Indeed, the difference between $q(t)$ and $x(t)$ is the error $e(t)$ incurred by the method and is shown in Figure \ref{fig:qssplot}b. It gets determined by the user-supplied accuracy-related parameters $\dQRel$ and  $\dQMin$, also shown in Figure \ref{fig:qssplot}c. If we consider for example the section starting at time $t_k$, we can see that both $q(t)$ and $x(t)$ evolve until the difference between them reaches $\deltaQ$. At that time, $q(t)$ is updated by quantizing $x(t)$, giving rise to a new section in the plot. This change is propagated to the ODE system by evaluating those state variables whose right-hand side depends on this variable. The coefficients of the polynomial approximations of $q(t)$ and $x(t)$ are presented in Figures \ref{fig:qssplot}d, \ref{fig:qssplot}e and \ref{fig:qssplot}f.

QSS methods can simulate any ODE system in the form of Equation \ref{eq:ODE} offering, among others, the following properties \cite{Kof01j}:

\begin{itemize}
    \item They are intrinsically asynchronous: 
    each state variable updates its value independently at self clocked time instants dictated by its own dynamics and the accuracy \deltaQ~(cf.\ time slicing methods where all state variables are scanned synchronously at each $\Delta t$). %This can offer significant performance advantages e.g. when simulating sparse systems.
    
    \item They provide dense output by means of the piecewise polynomial approximations of the state variables.
    
    \item They are very efficient at simulating systems with frequent discontinuities. A discontinuity is modeled by a zero-crossing function expressed in terms of the QSS polynomials. Thus, detecting a discontinuity calls only for finding the roots of a polynomial, which is computationally inexpensive for at most third-order QSS methods.% as it does not require iterative approximations.
    
    \item They satisfy strong stability and error bound theoretical properties.
    %QSS1 to QSS4 provide a global error bound (controlled by \deltaQ), limiting globally the error of the numerical solution of an analytically stable, time-invariant system.
\end{itemize}

Quantized State Systems cannot be formalized with the classical languages of differential nor difference equations. Rather, QSS can be exactly modeled and simulated by the Discrete Event System Specification (DEVS) \cite{Kof01j}, a formalism for modeling and simulation of generalized discrete event systems \cite{Zeigler}. Generalizing, we can say that QSS is a particular class of DEVS systems.
     
\subsubsection{QSS performance overview}
\label{sec:method_comparison}

QSS was extensively studied in several application domains by establishing a performance comparison against different discrete-time solvers. As an example, \cite{GAK12} compares QSS3 with the Runge-Kutta-Fehlberg and Bulirsch-Stoer methods (both enhanced with discontinuity handling routines) in the simulation of networks of spiking neurons. The authors found that QSS can offer significant performance improvements due to the efficient discontinuity handling and the activity-driven features. On the other hand, \cite{QSSPerf2} studies the simulation performance of one-dimensional advection-diffusion-reaction models comparing QSS against three  classic time discretization algorithms (one of them being DOPRI). It is shown that, in advection-–reaction-dominated situations, the second-order linearly implicit QSS method (LIQSS2) \cite{MK13} can yield performance improvements of at least one order of magnitude. LIQSS methods were also studied in the simulation of switched mode power supplies \cite{Mig15} and in the field of building performance simulation \cite{Ber18}. Models in these scenarios are typically stiff and present frequent discontinuities. In the former, the authors showed that LIQSS methods can be 3-200 times faster than the DASSL solver \cite{DASSL}. As for the latter, LIQSS2 was compared against different classic numerical solvers (including DASSL and DOPRI), achieving simulation speedups of at least one order of magnitude.

\subsection{Simulation toolkits}
\label{sec:toolkits}

\subsubsection{QSS Solver}
\label{sec:solver}

Traditionally, most implementations of QSS methods were provided by general-purpose discrete event simulation engines such as PowerDEVS \cite{BK11}. This generality usually brings about unnecessary CPU overheads (due to the underlying message-passing and/or event scheduling mechanisms) when the systems to be simulated are primarily continuous. QSS Solver, on the other hand, is an open-source standalone software that offers optimized implementations of the whole family of QSS methods, improving in more than one order of magnitude the computation times of previous discrete event implementations \cite{FK14}. It is composed by a set of modules implemented in the C programming language. One of these modules is a modeling front-end that allows the user to express  the models using $\mu$-Modelica \cite{bergero2012simulating}, a subset of the more general Modelica modeling language \cite{Modelica}. QSS Solver automatically generates C simulation code for any $\mu$-Modelica model. A simple graphical user interface integrates the solver engine with the modeling front-end and plotting and debug ancillary tools.

Modelica is a high-level, object-oriented language for modeling of complex systems. Models in Modelica are mathematically described by differential, algebraic and discrete equations. Sub–models can be inter–connected to create more complex models. Model composition can be done in a graphical way through several software tools.

\subsubsection{Geant4}
\label{sec:geant4}

\begin{figure}[h!]
\centerline{
    \includegraphics[scale=0.5]{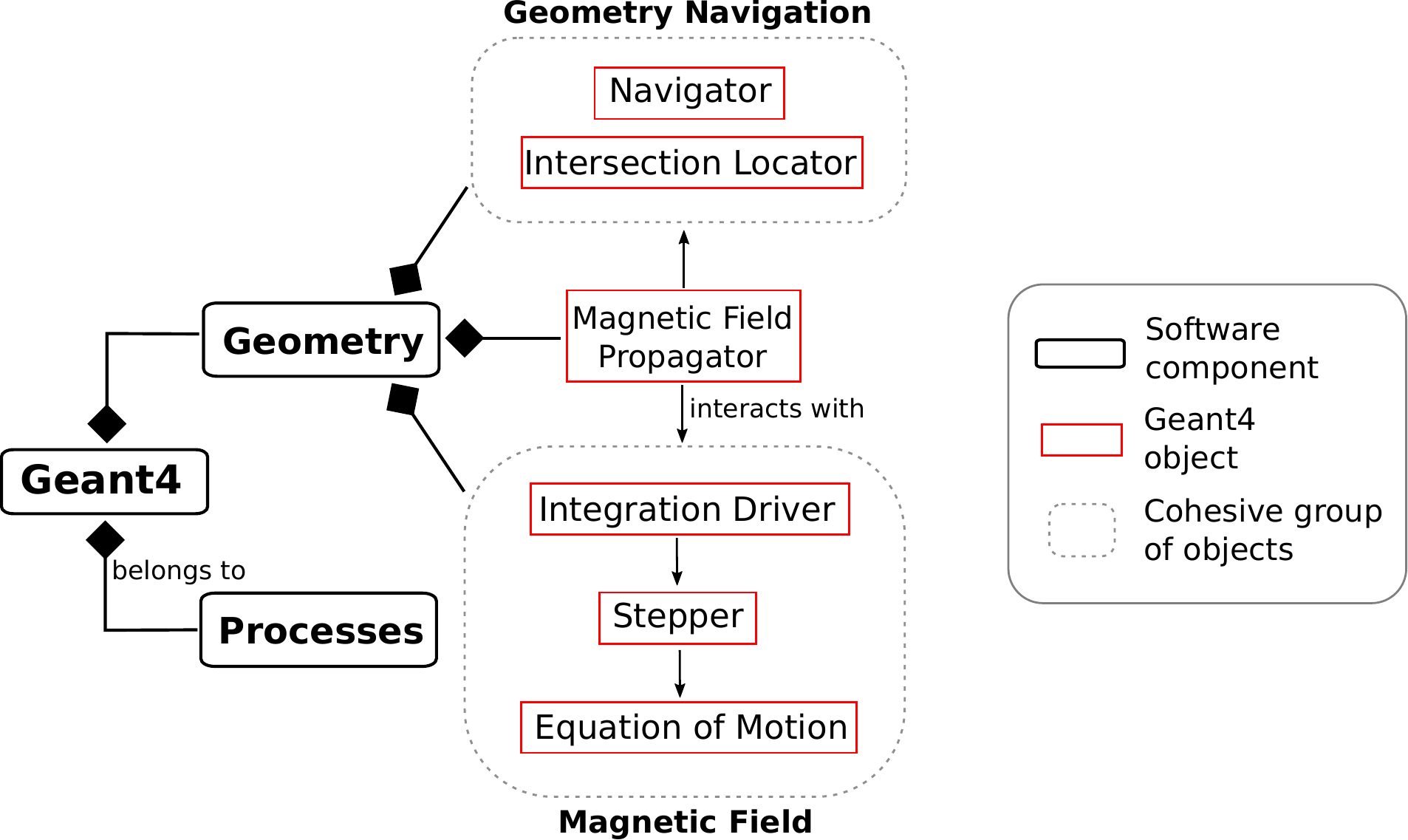}
}
    \caption{High-level structure of Geant4}
    \label{fig:G4}
\end{figure}

Geant4 is a software toolkit for the simulation of the passage of particles through matter that has been adopted as the simulation engine of choice for modern HEP experiments. It is composed of several \textsl{class categories} or \textsl{software components}, each one representing a cohesive set of classes with specific functionalities. Among them we highlight the \textbf{Geometry} component, which describes a geometrical structure (and the propagation of particles within it), and the \textbf{Processes} component, in charge of implementing the models of physical interactions. This is shown in Figure \ref{fig:G4}, which presents a simplified, high-level block diagram of Geant4. For further details please refer to the Geant4 documentation \cite{G4Guide} (Section \textsl{Class Categories and Domains}).

A Geant4 simulation is typically encompassed within a \textsl{run}, which consists of a series of \textsl{events}, the basic unit of simulation. An event is composed of one or more \textsl{tracks}, a snapshot of a particle at a particular point along its path, containing physical quantities such as energy and momentum. When a new event starts, \textsl{primary tracks} are generated and pushed into a simulation stack. Tracks are then popped up one at a time and simulated. Physical interactions might generate new, \textsl{secondary tracks}, which are also pushed into the stack and simulated serially (particles do not interact with each other). An event finishes when the stack becomes empty.

The trajectories of charged particles are computed through a complex algorithm detailed in the Geant4 documentation \cite{G4Guide} (Section \textsl{An Overview of Propagation in a Field}). This algorithm involves several accuracy parameters that can be supplied by the user: $\epsilon$ (controlling the relative error in the position and momentum), \deltaChord~(the maximum allowed distance of linear chord segments to the curved trajectory path) and \deltaIntersection~(which imposes accuracy constraints to the calculation of volume boundary crossings). The \textsl{Magnetic Field Propagator} object orchestrates the propagation of charged particles in a magnetic field by means of this algorithm. A trajectory is made up of \textsl{steps} that advance the particle a given distance (which can be limited to a maximum fixed size by the user through the \stepMax~parameter). Each of these steps are computed by the \textsl{steppers}, which provide custom implementations of Runge-Kutta-based numerical integration algorithms to solve the underlying equations of motion. The Magnetic Field Propagator, however, does not interact directly with the steppers but with an \textsl{Integration Driver} that exposes a common integration interface for different kinds of numerical methods.

A step might end before covering its length due to reaching a volume boundary. When Geant4 detects a boundary crossing, it executes an iterative algorithm to compute the intersection point within certain accuracy constraints. This is implemented by the \textsl{Intersection Locator} object.

\begin{figure}[h!]
\centerline{
    \includegraphics[scale=0.5]{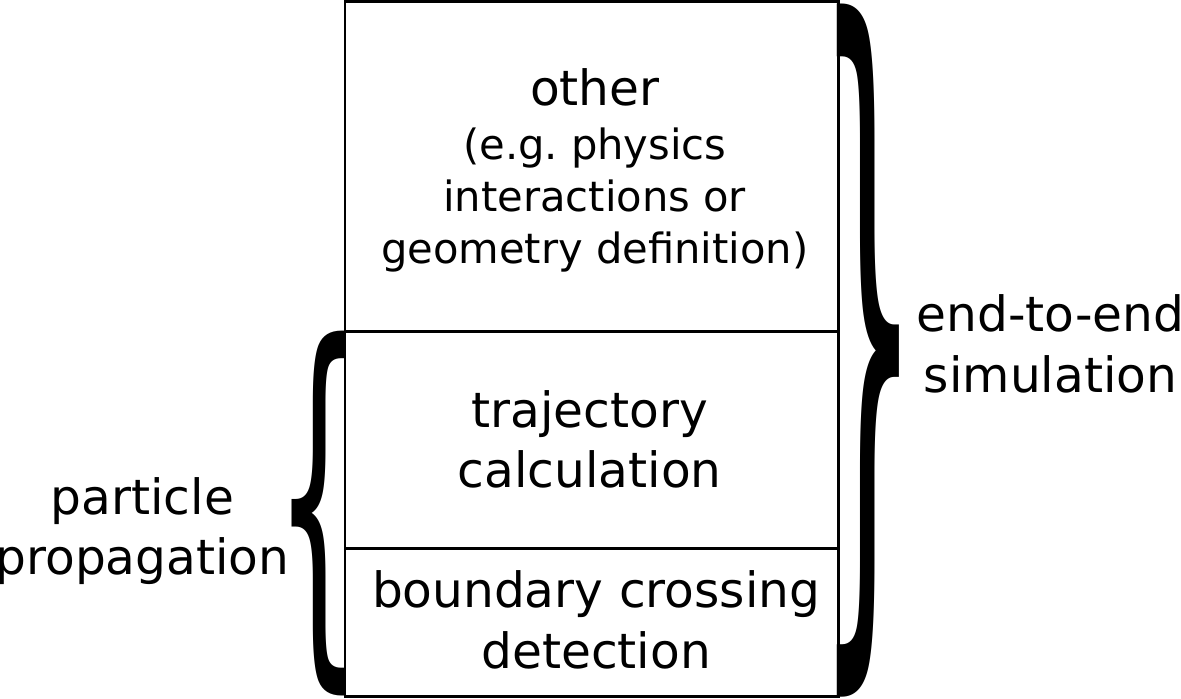}}
    \caption{Main components of a Geant4 simulation}
    \label{fig:stepping_div}
\end{figure}

The task of propagating charged particles can be thus decomposed into a \textsl{trajectory calculation} part and a \textsl{boundary crossing detection} part \cite{agostinelli2003geant4}. This is shown in Figure \ref{fig:stepping_div}. Our research is focused on this particle propagation component of complete, end-to-end Geant4 simulations, leaving out of scope other aspects such as physics evaluation.

Since version 10.4, released in December 2017, the default stepper in Geant4 is the fifth/fourth-order accurate Dormand-Prince adaptive method (\DOPRI). Previous versions used the classical fourth-order Runge-Kutta (\RK) as default.

\subsection{Related work}
\label{sec:related}

There are several other particle simulators used by the HEP community aside from Geant4. For example, MARS15 \cite{MARS} is a set of Monte Carlo programs for the simulation of hadronic and electromagnetic
cascades in 3D geometries. MCNP \cite{MCNP} is another Monte Carlo-based software for simulating nuclear processes, whereas FLUKA \cite{FLUKA} is a general-purpose Fortran simulation package for particle transport and interactions with matter. Closely related to Geant4 is the GeantV project \cite{GeantV2020}. This prototype (recently discontinued) aimed at re-engineering Geant4 to run on modern, parallel computing architectures. Among these alternatives, we based our research on Geant4 as it is currently the simulation engine of choice for most HEP experiments \cite{HEPChallenges}.

As for related co-simulation techniques, the modern Functional Mockup Interface (FMI) \cite{FMI} allows for the concurrent simulation of subsystems with certain synchronization points where data is exchanged. The FMI standard is considered the most promising standard for continuous time,
discrete event and hybrid co-simulation, as discussed in a recent empirical survey on co-simulation \cite{CosimSurvey}. The co-simulation software MpCCI \cite{Wirth2017} provides algorithms and interfaces to couple different tools so as to solve collaboratively the simulation of disjoint models. Another strategy parallel by design is Complex Control Systems Simulation (CCSS) \cite{Munawar2013}, which is a distributed discrete event co-simulation technique with application to the automotive industry.

We studied the performance of QSS Solver against Geant4 for the simulation of a baseline HEP setup  \cite{San16} (a charged particle in a uniform magnetic field describing a circular 2D motion crossing parallel planes) which for the sake of context will be reviewed in Section \ref{sec:chep}. This work was continued by designing and implementing a proof-of-concept version of GQLink \cite{San17}, conceived as an abstract interface that allows connecting Geant4 to arbitrary external integrators, in particular the QSS family as implemented by QSS Solver. Finally, in \cite{San18} we formalized the underlying co-simulation concepts and developed several algorithmic and low-level optimizations to improve GQLink's simulation performance. In this work we perform additional and more comprehensive performance studies in the context of a new Geant4 version, including the new de-facto Geant4 stepper in the comparisons.

%%%%% MOTIVATING CASE STUDIES %%%%%%%%%%%
\section{Motivating case studies}
\label{sec:casestudies}

We start by describing the two main HEP setups that will be used throughout our work to establish the aforementioned performance comparisons.

\subsection{Oscillating particle under a constant magnetic field}
\label{sec:helix}

Our first scenario serves as an introductory baseline test case engineered to shed light on the core aspects of a given integrator: its error control capabilities, its step computation performance and its efficiency to detect boundary crossings. In order to fulfill this purpose, the most relevant features of this case study are the following:
\begin{itemize}
    \item It offers a closed-form analytic solution which makes the error analysis accurate and straightforward.
    \item Particle-matter interactions are explicitly turned off so that there is no CPU overhead involved in physics computation.
    \item The number of boundary crossings can be easily controlled by the user through a set of runtime parameters.
\end{itemize}

We introduce two variants which are loosely based on the basic example B2 that is shipped along with the Geant4 source code \cite{G4Release1005}. This example focuses on the most typical use-cases of a Geant4 application.

\begin{figure}[hbt]
\begin{center}
\begin{tabular}{cc}
  \subfloat[2D setup\label{fig:circle}]
    {
        \includegraphics[scale=0.4,valign=t]{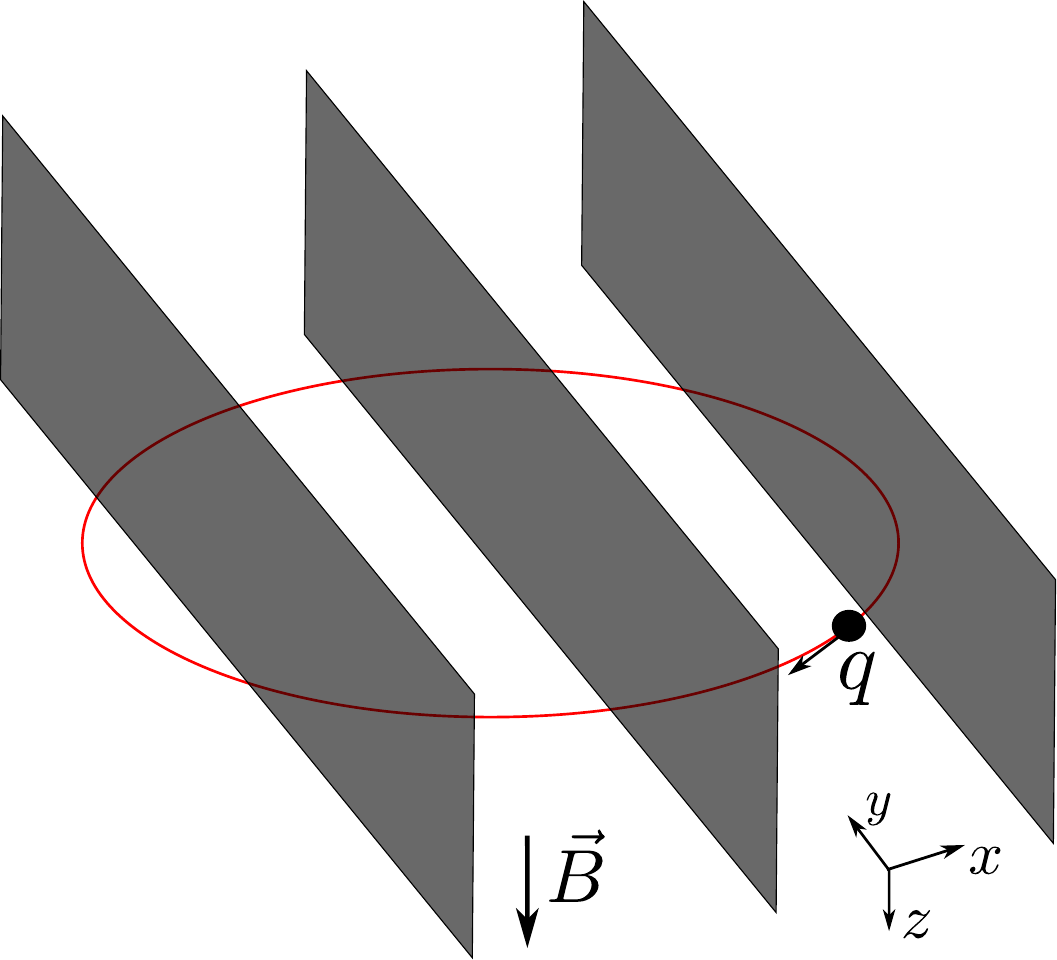}
    } &
    
  \subfloat[Sample trajectories inside a 3D cube mesh\label{fig:helix}]
    {
        \includegraphics[scale=0.17,valign=c]{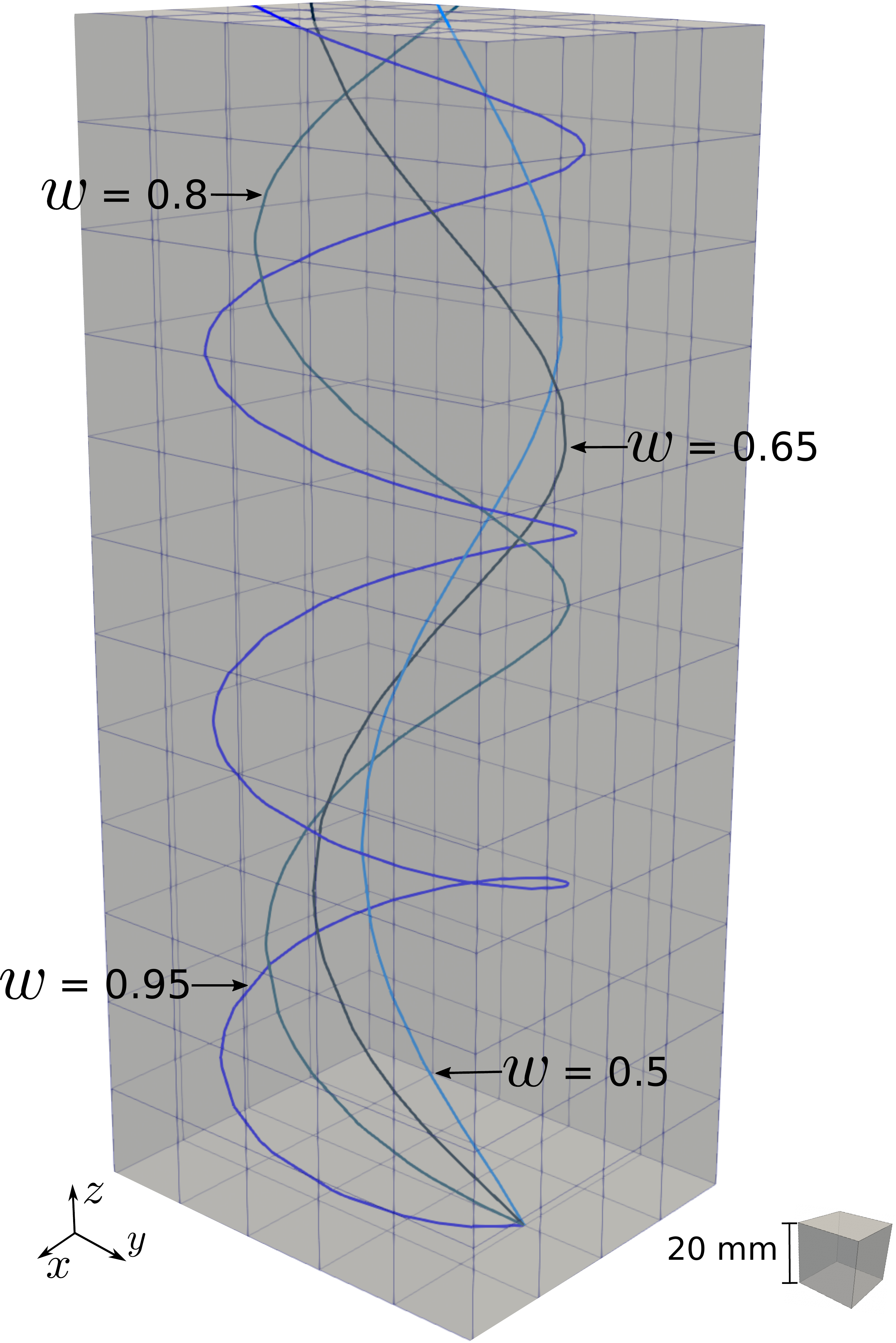}
    }
\end{tabular}
\caption{Baseline case study sketches}
\label{fig:n02}
\end{center}
\end{figure}

\paragraph{2D setup: B2c} Its most simple implementation consists in a single positron under a uniform, static magnetic field along the $\hat{z}$ plane, i.e., $\vec{B}=(0,0,B)=B \hat{z}$ with initial velocity $\vec{v} = v\hat{x}$, where $v = 0.999 \, c$ and $c = 299.792458 \, \textrm{mm}/\textrm{ns}$ is the speed of light. Thus, the particle describes a circular trajectory in the $(\hat{x},\hat{y})$ plane. The magnetic field density $B$, measured in teslas, can be supplied as a model parameter.
Boundary crossings are modeled by inserting equidistant parallel planes along the trajectory of the particle. This is shown in Figure~\ref{fig:circle}. The user can control the number of boundary crossings by specifying how many planes should be inserted.

\paragraph{3D extension: B2h} We also developed a 3D extension of this setup in which the particle follows a helical trajectory with a linear increase in $\hat{z}$ with respect to time. This is achieved by generalizing the initial velocity, which is now $\vec{v} = (w \cdot v\hat{x}, 0, \sqrt{1 - w^2} \cdot v\hat{z})$. The coefficient $w \in [0,1]$ is a model parameter that controls the initial speed in $\hat{x}$. In this extension, the geometry is a 3D cube mesh where the cube edge length can be supplied by the user. Thus, as the cube size decreases, the number of boundary crossings along the trajectory of the particle increases. Note that this scenario enables boundary crossings in each of the three dimensions and to different angles, as opposed to its 2D counterpart. Figure~\ref{fig:helix} shows a 3D mesh of cubes, each of which with a cube edge length of $20~\textrm{mm}$. Four trajectories produced with different values of $w$ appear inside it (the magnetic field density $B$ is set to 1 T).

\subsection{A real-world particle detector: CMS}
\label{sec:cms}

The primary goal of this case study is to analyze how our particle propagation strategies perform in a realistic HEP setup featuring a complex magnetic field, a large geometry composed of different types of volumes and with physics interactions enabled. In this sense, this case study complements the baseline tracking experiment introduced in Section~\ref{sec:helix}.

To that end we simulate the Compact Muon Solenoid (CMS) particle detector. The CMS experiment is a general-purpose particle detector on the Large Hadron Collider (LHC) at CERN. It was designed to study a broad range of physics, spanning the exploration of physics at high energies, searching for evidence of extra dimensions and testing proposed theories of dark matter. 

We adopted \texttt{cmsExp}, a \textsl{third-party}  standalone Geant4 application proposed in \cite{Dotti2015}. It was originally conceived as a benchmarking asset for Geant4 aimed at measuring its performance in a typical, complex HEP detector. The application uses the detector's full Run 1 geometry (i.e., from the LHC first operational run, which took place between 2009 and 2013) and a volume-based magnetic field taken from CMS Offline Software (\url{https://github.com/cms-sw/cmssw}). Essentially, given a point in space, the value of the magnetic field is obtained by interpolation from a regular grid of values \cite{VolumeBasedField}. As for the physics, we use the recommended reference physics list \texttt{FTFP\_BERT} \cite{G4Phys2011}.

It is important to note that, as opposed to case study \ref{sec:helix}, there is no closed-form analytical solution for CMS simulations. Nevertheless, we validate our simulations by checking their consistency by means of statistical tests (further explained in Section \ref{sec:stepper_results_cms_validation}).

A CMS simulation starts with a primary particle being injected into the detector through a \textsl{particle gun}, which is configured by the user to shoot a given particle with an initial kinetic energy and momentum direction. As the primary particle interacts with matter along its trajectory, several secondary particles are generated and eventually simulated, which in turn can lead to additional secondaries being simulated. Over all, in the typical CMS simulations we ran, there are more than $62000$ secondary particles per event.

\begin{figure}[h!]
\centerline{
    \includegraphics[scale=0.3]{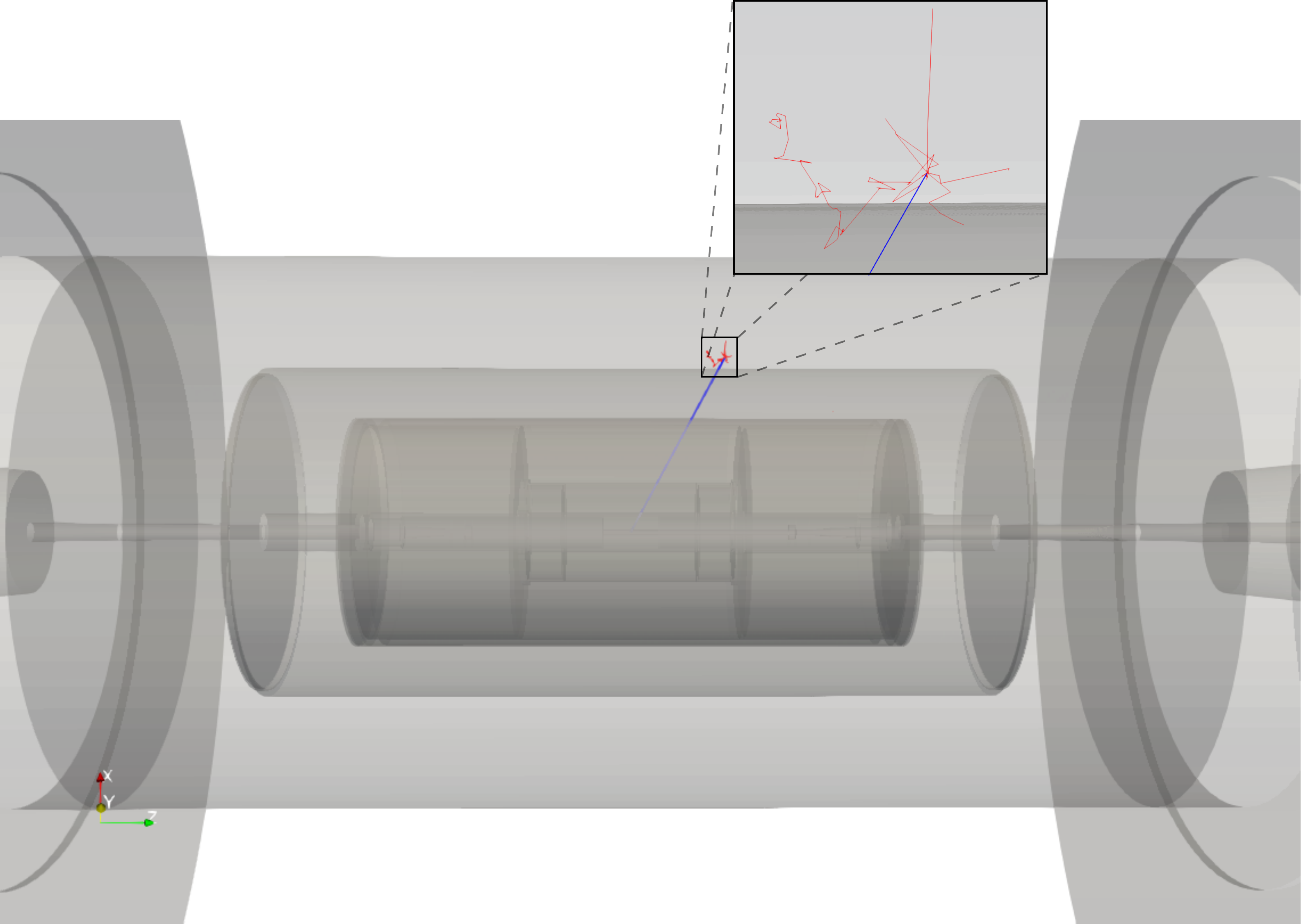}
}
    \caption{A section of the CMS geometry with 20 particle trajectories}
    \label{fig:cms}
\end{figure}

Figure~\ref{fig:cms} shows a zoomed-in section of the CMS geometry along with the first 20 trajectories (primary particle in blue) extracted from a single simulation. 

%%%%%%%%%%%% QSS IN HEP %%%%%%%%%%%%%%%%
\section{Feasibility study of QSS for HEP applications}
\label{sec:chep}

Towards the goal of determining whether HEP applications can profit from discrete-event numerical solvers, a first essential step is to study their performance in a controlled, yet representative scenario that can properly stress their most fundamental features. We then designed a performance comparison between QSS Solver and Geant4 in a tracking-intensive setup. The focus is made on accuracy and integration efficiency, the most relevant performance capabilities of the solvers at hand. In this Section we will summarize the main results achieved by this study, which enables us to provide a reference framework to properly contextualize and motivate our further developments. 

\subsection{Scenario description}
\label{sec:chep_scenario}

The proposed test scenario is case study B2c (Section \ref{sec:helix}), where a positron describes a circular, 2D trajectory due to the action of an uniform magnetic field in $\hat{z}$ (Figure~\ref{fig:circle}). Parallel equidistant planes can be injected on demand in the geometry to exercise the boundary crossing resolution capabilities of the numerical solver. This key feature enables to test whether QSS can leverage its efficient discontinuity handling properties in the context of a typical HEP setup. Moreover, a deterministic, rigorous accuracy comparison can be carried out due to the existence of a closed-form analytical solution, which is in turn a consequence of the intended simplicity of the model.

\subsection{Implementation}
\label{sec:chep_impl}

For Geant4, we developed a standalone tracking application with physics interactions disabled. The geometry is constructed after a user-configurable parameter that specifies the number of crossing planes. The chosen stepper is the vastly used fourth-order Runge-Kutta (\RK), the default stepper for Geant4 until it was replaced in release 10.4 of December 2017 (see Section 4 of the Geant4 10.4 release notes \cite{G4ReleaseNotes} for further details). The underlying equations of motion are the usual Lorentz equations for charged particles in a magnetic field (Equation \ref{eq:lorentz}), provided by default in Geant4. There, $q$ and $m$ stand for the charge and mass of the particle, respectively; $c$ is the speed of light; $\gamma$ is the Lorentz factor and $\vec{B}$ is the magnetic field.

\begin{equation}
\systeme{\dot{x} = v_x \quad\quad \dot{v_x} = \frac{q \, c^2}{m \, \gamma} \cdot (v_y \, B_z - v_z \, B_y),
          \dot{y} = v_y \quad\quad \dot{v_y} = \frac{q \, c^2}{m \, \gamma} \cdot (v_z \, B_x - v_x \, B_z),
          \dot{z} = v_z \quad\quad \dot{v_z} = \frac{q \, c^2}{m \, \gamma} \cdot (v_x \, B_y - v_y \, B_x)
\label{eq:lorentz}}    
\end{equation}

\begin{figure}[h!]
\centerline{
    \includegraphics[scale=0.8]{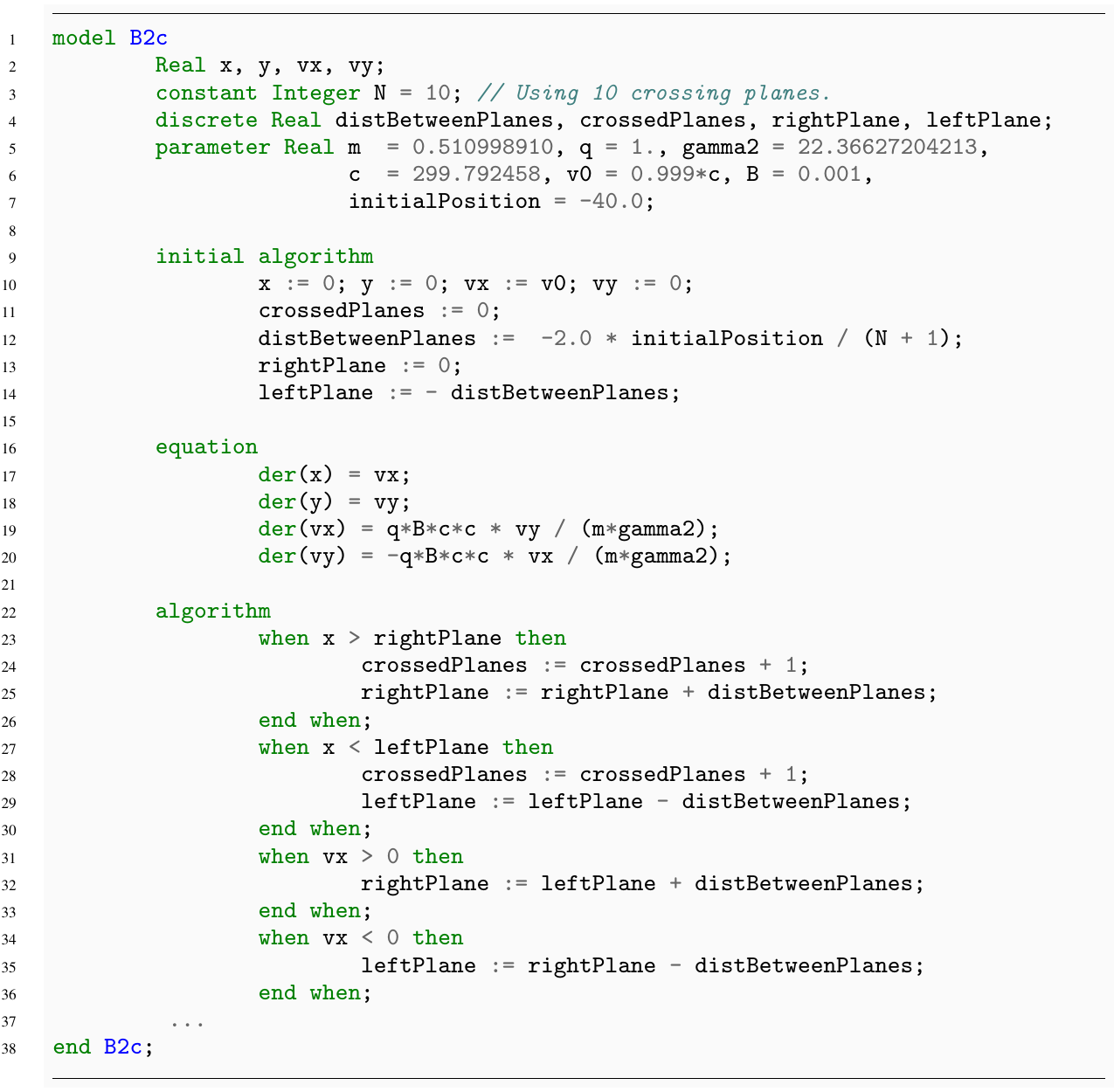}
}
    \caption{$\mu$-Modelica model for case study B2c}
    \label{fig:b2c_model}
\end{figure}

As for QSS Solver, we formulated a $\mu$-Modelica model that provides an implementation of the Lorentz equations. In this model, crossing planes are coded in the form of discrete events, generated whenever the particle trajectory passes through the predetermined coordinates of each plane. Figure \ref{fig:b2c_model} shows the complete model with 10 crossing planes. We chose third-order QSS (QSS3) as the integrator to perform the  experimentations.

Accuracy parameters \deltaChord~and \deltaIntersection~in Geant4 were set to 0.25 mm and $10^{-5}$ mm, respectively. We set the magnetic field density $B$ to 1 tesla and
defined a track length of 1 km. This very long track length allows for stressing the error control capabilities of the underlying numerical methods, an essential goal of this study. The error in the propagation of charged particles in uniform magnetic fields (using standard numerical methods such as \RK) can be very small for typical distances (in the order of millimeters or centimeters) of HEP detector simulations. For the proposed track length of 1 km, we found a maximum absolute error of about 3 mm for \RK, i.e., a 0.0003\% of the track length. Please refer to \cite{San16} for further details about this study.

\subsection{Performance comparison}
\label{sec:chep_performance}

\begin{figure}[hbt]
\begin{center}
\begin{tabular}{cc}
  \hspace{-1cm}
  \subfloat[Maximum error (left $y$ axis, dashed lines) and simulation time (right $y$ axis, solid lines) vs. relative precision\label{fig:chep_noplanes}]
    {
        \includegraphics[scale=0.3,valign=t]{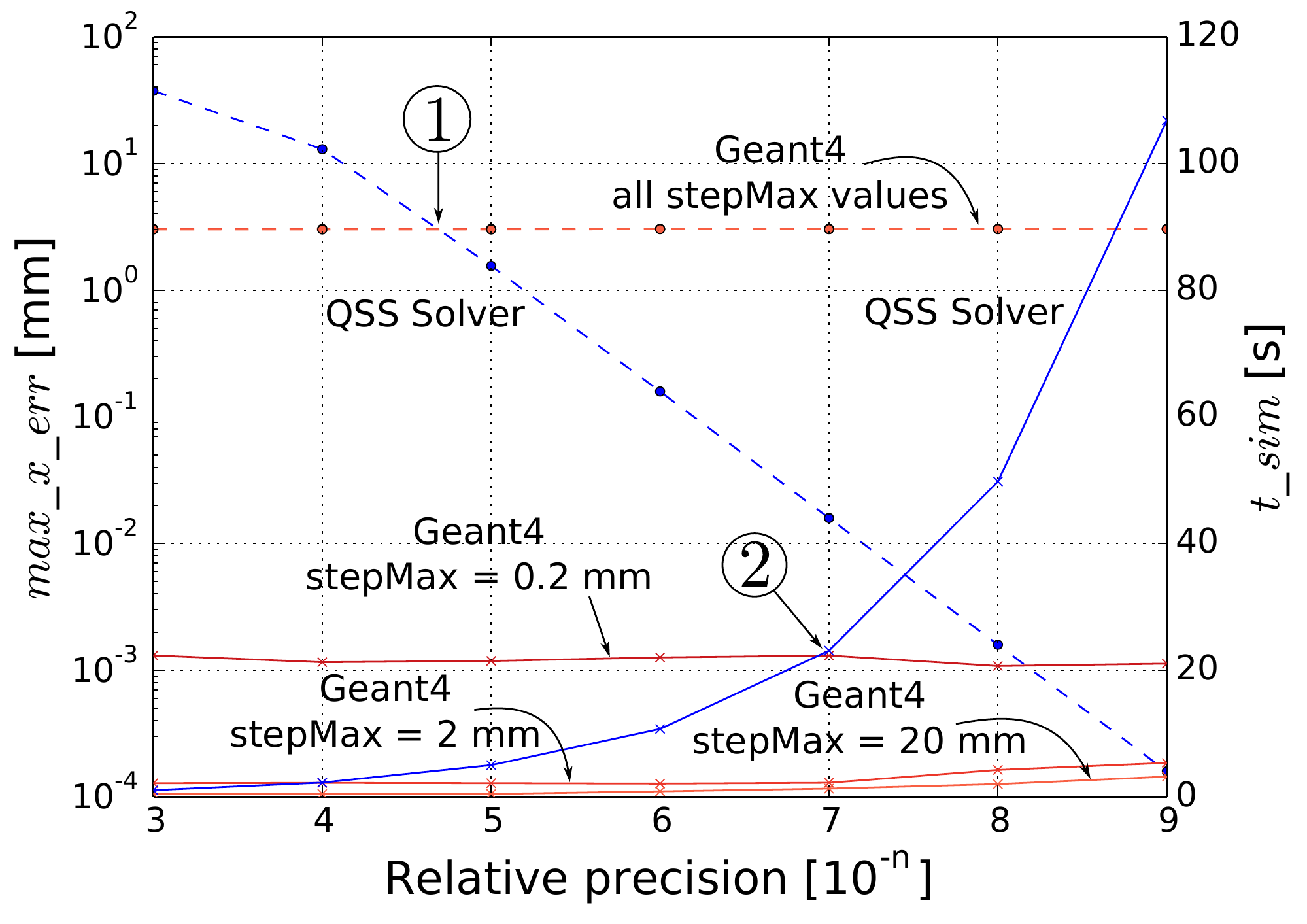}
    } &
    
  \subfloat[Maximum error (left $y$ axis, dashed lines) and simulation time (right $y$ axis, solid lines) vs. number of crossing planes\label{fig:chep_planes}]
    {
        \includegraphics[scale=0.35,valign=t]{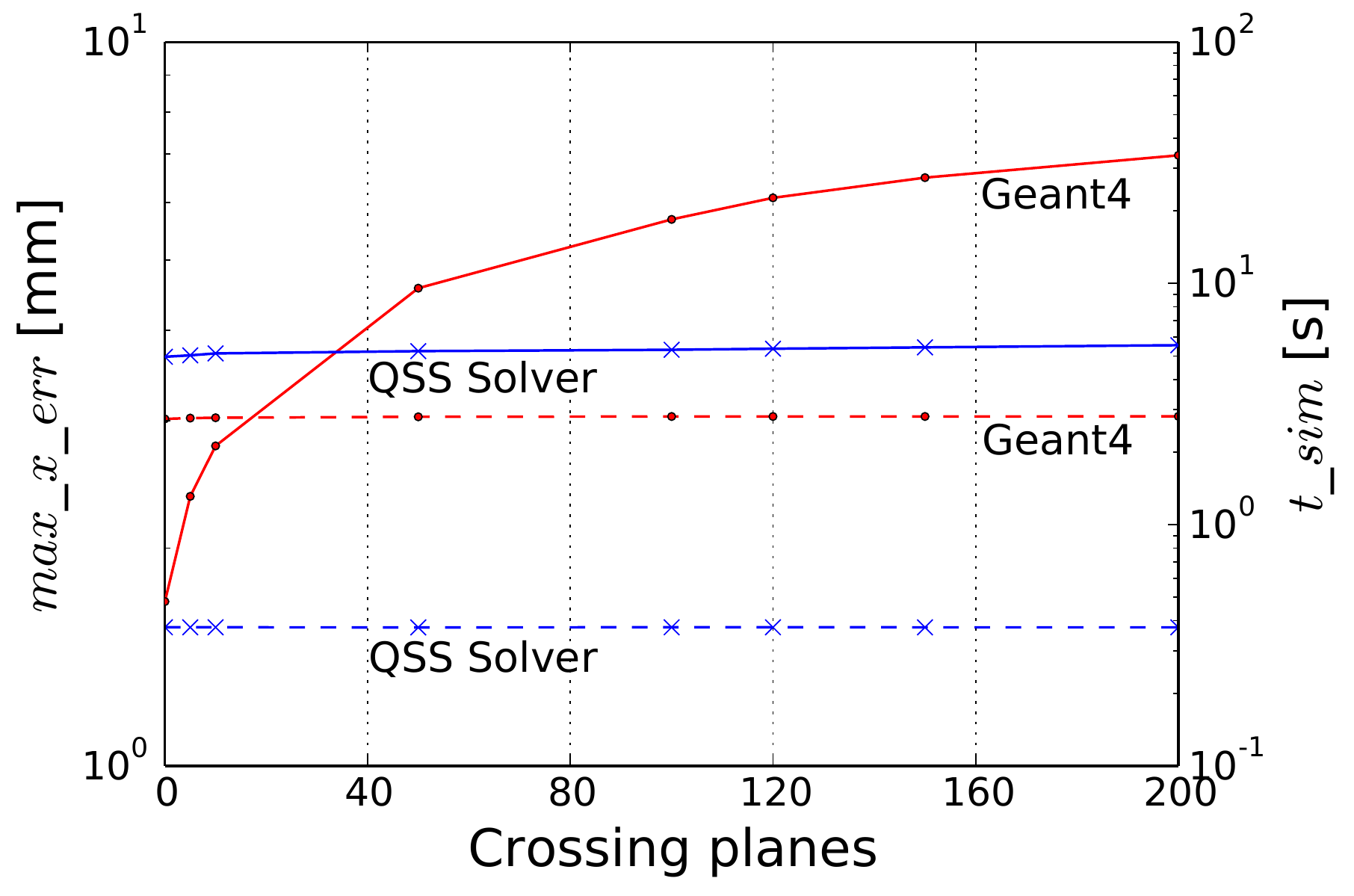}
    }
\end{tabular}
\caption{Performance comparison between QSS Solver and Geant4}
\end{center}
\end{figure}

Our first analysis studied the performance of both toolkits when there are no crossing planes in the geometry. For this purpose, we experimented with three values of \stepMax~in Geant4 (0.2 mm, 2 mm and 20 mm) and we swept a range of values for relative accuracy ($\epsilon$ parameter in Geant4; $\dQRel$ in QSS Solver) in order to compare the numerical error and the performance of each simulation. The error, termed $max\_x\_err$, was calculated as the maximum absolute difference observed with respect to the analytic solution along the $x$ coordinate of the particle's position. We measured the end-to-end simulation time $t\_sim$ to establish a performance comparison. Results are shown in Figure~\ref{fig:chep_noplanes}. First, we can see that \stepMax~ does not seem to have a perceivable impact on the error, as the respective dashed lines appear to be on top of each other. Moreover, the Geant4 application does not seem to improve the error when the relative precision is increased (i.e., when $\epsilon$ is decreased). This may be due to a combination of the higher order of the chosen stepper (\RK) with other accuracy parameters (in particular, \deltaChord), which remained fixed for every point in the curves. On the other hand, as indicated by \encircle{1}, QSS3's error decreases approximately by an order of magnitude for each extra order of magnitude of $\dQRel$. We also note that lower values of \stepMax~yield higher simulation times in Geant4, which is expected since they demand a higher number of steps to cover the same track length. For QSS3, we observe that the simulation time increases with the cubic root of $\dQRel$, which confirms a theoretical property of the method \cite{Kof06a}. When \stepMax \, $= 0.2$~mm, between points \encircle{1} and \encircle{2}, QSS3 outperforms Geant4 achieving both smaller error bounds and lower simulation times.

Both methods were also studied in the context of and increasing number of crossing planes along the trajectory of the particle, using a fixed relative accuracy of $10^{-5}$. This study is depicted in Figure~\ref{fig:chep_planes}. We found that QSS Solver's QSS3 simulation time scales better than Geant4's \RK as the number of plane increases, observing performance improvements of at most 6x for 200 planes. The smooth growth in QSS3's simulation time is a direct consequence of the key QSS feature of efficient discontinuity handling. Finally, we observed that the error is not significantly affected by the number of boundary crossings in both toolkits. QSS Solver achieved better error bounds than Geant4.\\

This analysis suggests that QSS Solver can offer performance improvements for HEP simulations featuring intense volume crossing activity. Yet, there are salient differences between the simulation toolkits impeding direct and fair comparisons. Moreover, QSS Solver alone cannot simulate complex HEP setups. However, the outcome provides a solid foundation that justifies the efforts towards an integration of QSS with the Geant4 transportation engine. We will accomplish this progressively, in two steps built upon each other: a co-simulation strategy that connects the QSS Solver simulation engine with Geant4, and a new embedded QSS stepper developed natively for the Geant4 particle transportation ecosystem.

%%%%%%%%%%%%%%% GQLINK %%%%%%%%%%%%%%%%%%%
\section{First approach: A co-simulation strategy}
\label{sec:cosimulation}

In this approach we connect the engine of QSS Solver with Geant4 in such a way that the responsibilities of step computation can be transparently delegated. This interconnection serves multiple purposes. First, it enables to test whether Geant4 and QSS can be smoothly and correctly coupled. Besides, it leverages the vast domain-specific knowledge on physics interaction already available in Geant4. Also leveraged is the optimized  state-of-the-art implementation of QSS methods in the QSS Solver engine. In turn, the development overhead is minimized, as it is not needed to engineer and design a QSS implementation specially tailored for Geant4. In fact, this approach demands at most devising a clean and sound method for connecting the toolkits. Finally, by delegating the particle propagation responsibilities to QSS Solver, other geometry navigation algorithms can be tested and explored. One such example is based on approximating the geometry volumes by faceted polyhedrons. Thus, intersection points can be computed more efficiently by solving a polynomial equation involving these facets and the QSS polynomials that approximate the particle trajectory.

With these goals in mind we developed GQLink (\textsl{Geant4-to-QSS Solver Link}), a collaborative co-simulation strategy where Geant4 drives the simulation and delegates particle transport to QSS Solver. Thus, the former is in charge of defining and evaluating the physics processes, whereas the latter is responsible for integrating the equations of motion and transporting particles, querying the geometry navigation routines provided by Geant4 to detect volume crossings. 

\subsection{High-level architecture}
\label{sec:gqlink_arch}

\begin{figure}[h!]
\centerline{
    \includegraphics[scale=0.5]{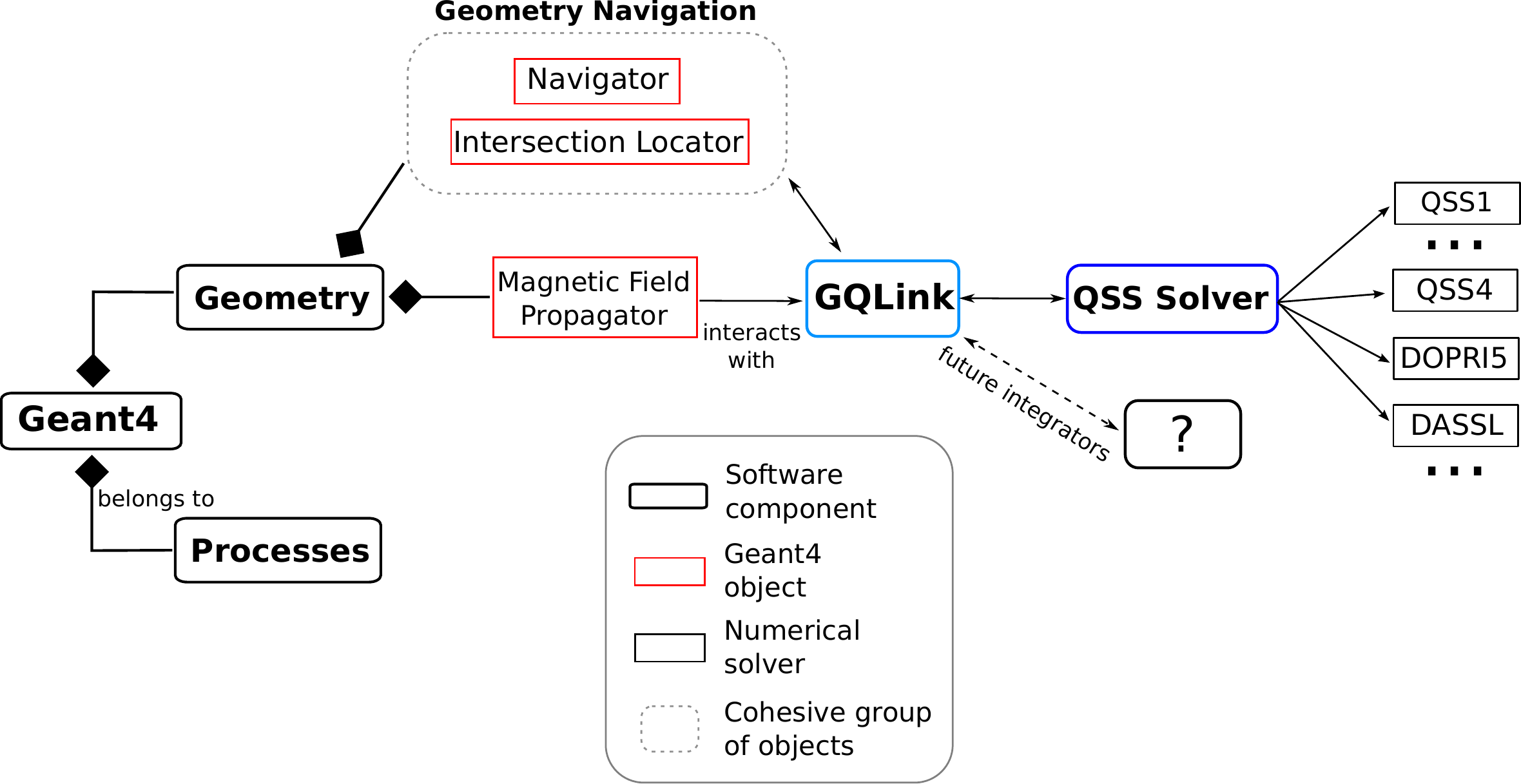}
}
    \caption{High-level structure of Geant4 with GQLink}
    \label{fig:GQLink}
\end{figure}

Figure \ref{fig:GQLink} shows a high-level architecture where  GQLink takes over the particle tracking actions typically in control of the Magnetic Field-related objects in the \textbf{Geometry} software component (cf. Figure \ref{fig:G4}). As such, the Magnetic Field Propagator no longer interacts with the standard Geant4 integration drivers and steppers when GQLink is enabled. In order to complete an integration step, GQLink relies upon QSS Solver, the chosen underlying integration engine, to solve the equations of motion. However, one of the design goals of GQLink is to stand for an abstract, transparent particle propagation interface that can be eventually used to connect Geant4 with other external solvers. 

While a step is being computed, GQLink periodically queries the geometry navigation objects of Geant4 to properly detect volume boundary crossings. These are bidirectional interactions, as the intersection finding algorithm sends a query back to GQLink on each iteration in order to improve its estimation of the intersection point. This will be further elaborated in Section \ref{sec:gqlink_cosim}.

\subsection{Model definition}
\label{sec:gqlink_model}

While Geant4 gets initialized, before running a simulation, a set of user-supplied parameters is used to choose the GQLink model to simulate. So far, we tested standard equations of motion of charged particles in a magnetic field (Equation \ref{eq:lorentz}), which are available in a set of $\mu$-Modelica models that are pre-compiled into shared libraries. Each of these shared libraries correspond to different QSS methods (e.g., QSS1 to QSS4). Thus, Geant4 picks the one satisfying the user requirements, dynamically links to it and initializes a GQLink Model object. This object exposes an API used by the Magnetic Field Propagator to interact with QSS Solver.

Further models (e.g., involving electric fields or particle spin) can be added by expressing the equations in $\mu$-Modelica and putting the resulting file into the GQLink folder. A new rule added to the Geant4 build system compiles in two steps the files in this folder to produce the aforementioned shared libraries. First, the $\mu$-Modelica compiler generates C code from the model sources. This code is later compiled into a shared object by the usual C/C++ compiler.

\subsection{Co-simulation interactions}
\label{sec:gqlink_cosim}

\begin{figure}[h!]
\centerline{
    \includegraphics[scale=0.35]{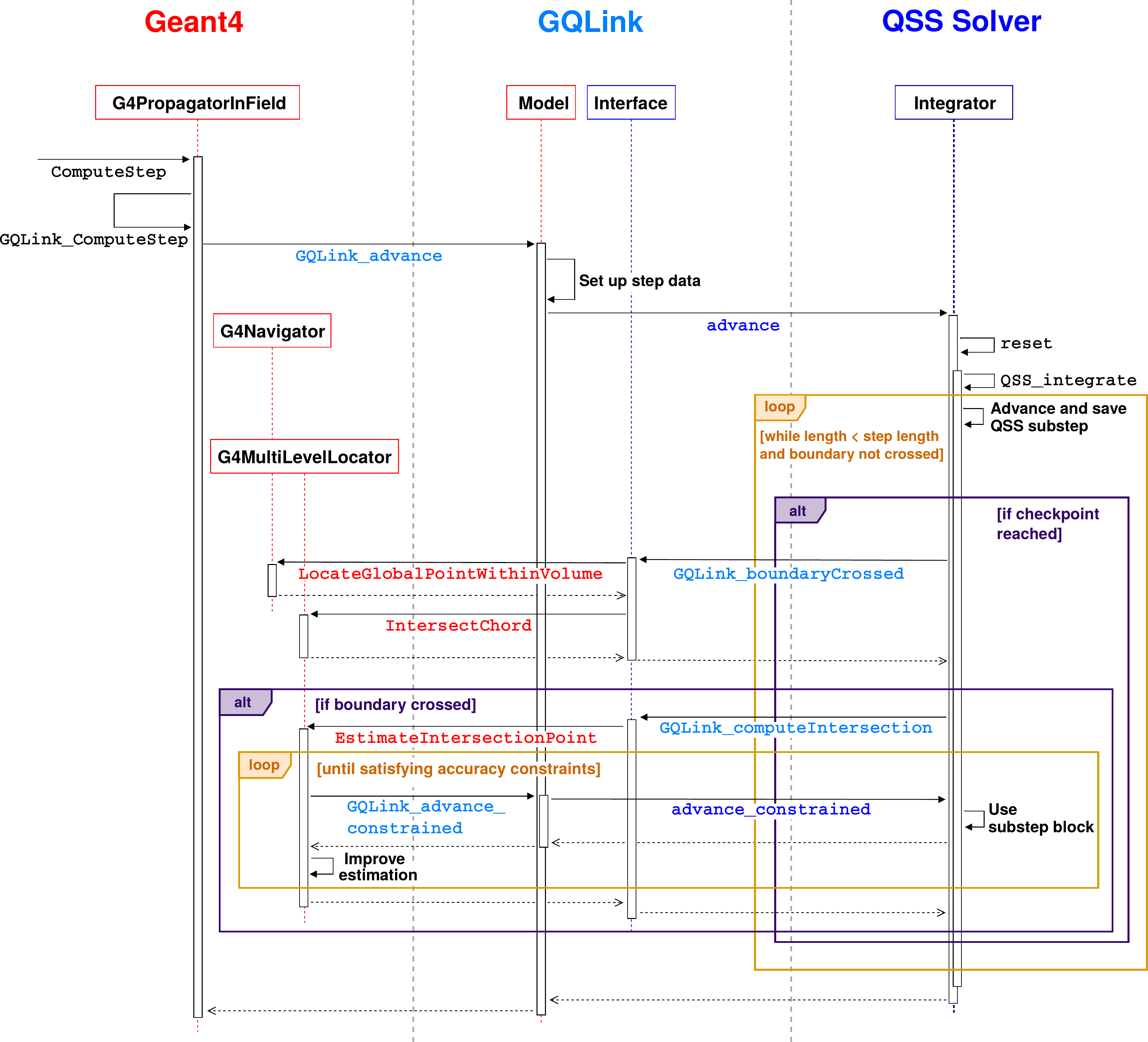}
}
    \caption{Co-simulation interactions in GQLink}
    \label{fig:GQLink_simulation}
\end{figure}

Figure \ref{fig:GQLink_simulation} illustrates the main co-simulation interactions between Geant4 and QSS Solver through GQLink during a step computation, whose entry point is the\\ \texttt{G4PropagatorInField::ComputeStep} method implemented by the Magnetic Field Propagator. GQLink is invoked after calling the new method \texttt{G4PropagatorInField ::GQLink\_ComputeStep}, which interacts with the GQLink Model object and instructs it to advance a given length. Once a data structure is initialized with assorted information relevant for the step (e.g., particle properties and step length), the advance request is relayed to the QSS Solver integration engine. It first resets the internal simulator state using the particle charge, velocity and position and current magnetic field value. This reinitialization is essential to ensure correctness, as the post-step physics processes evaluated by Geant4 might change some particle properties (e.g. its direction). Next, the main integration routine of the QSS engine takes over control. This routine will iterate as long as the requested step length is covered, or otherwise until a volume boundary is crossed. Each iteration gives rise to a so-called QSS \textsl{substep}, which can be understood as a tuple $\left<t, v, l, \mathbf{x}\right>$, where $t$ is its start time, $v$ the velocity of the particle along the substep, $l$ the traversed length upon completing the substep and $\textbf{x}$ the vector function that approximates the particle trajectory along the substep by means of polynomial functions. Substeps are packed into a \textsl{substep block} with a user-configurable size $n$. Once $n$ substeps are packed, the integration routine is suspended by a \textsl{checkpoint} where Geant4 is queried via GQLink for possible boundary crossings. This process follows the same routine call pattern implemented by Geant4's transportation engine: after locating the starting point of the substep block in the geometry (via method \texttt{G4Navigator::LocateGlobalPointWithinVolume}), a linear segment connecting the endpoints of the block is tested for volume intersections (via method \texttt{G4MultiLevelLocator::IntersectChord}). If a volume crossing is detected, the intersection point found is taken as an initial estimation of the actual boundary crossing point, which is computed by the method \texttt{G4MultiLevelLocator::Estimate\-IntersectionPoint}. This method keeps refining the successive approximations until the intersection accuracy constraints are met. On each iteration, QSS Solver is queried back to advance a certain length constrained to the current substep block. This is efficiently computed by finding first the appropriate substep and finally evaluating the corresponding QSS polynomials that approximate the particle trajectory.
For a more comprehensive explanation of the particle propagation interactions and the intersection finding algorithm, please refer to \cite{San18}.

\subsection{Changes made to Geant4}
\label{sec:gqlink_changes}

We leveraged Geant4's cohesive object-oriented design to implement GQLink with a very limited number of changes introduced to the source code. We summarize them below:
\begin{itemize}
	\item The Magnetic Field Propagator (\texttt{G4PropagatorInField} class) was extended with a pointer to the GQLink Model object and two new methods: one for initializing GQLink (following the actions detailed in Section \ref{sec:gqlink_model}) and the other for computing a step (\texttt{GQLink\_ComputeStep}, explained in Section \ref{sec:gqlink_cosim}).
	
	\item The Intersection Locator (\texttt{G4MultiLevelLocator} class) also holds a pointer to GQLink's Model. It is used to query QSS Solver on each iteration of \texttt{Estimate\-Intersection\-Point}, as explained in Section \ref{sec:gqlink_cosim}. Actually, this is done inside the method \texttt{ApproxCurvePointV} of class \texttt{G4ChordFinder} (the Model object is passed as an argument).
	
    \item Geant4's \texttt{cmake}-based build system was extended with a new rule to produce the shared libraries from the $\mu$-Modelica model sources, as described in Section \ref{sec:gqlink_model}. Also, upon building Geant4, new libraries for GQLink and the QSS Solver engine are compiled. The former is a new dependency for the \textbf{Geometry} component.
    
    \item We also included a new \texttt{cmake} option to enable or disable GQLink code upon compilation (\texttt{GEANT4\_USE\_GQLINK}).
\end{itemize}

\subsection{Simulation experiments and discussion}
\label{sec:gqlink_results}

In this Section we will study how GQLink performs against the two most relevant Geant4 steppers (\RK and \DOPRI) in the context of case study B2h (introduced in Section \ref{sec:helix}). The hardware and software platform used throughout the experimentation is described in \ref{sec:setup}. The underlying dataset for this analysis can be found in \cite{ExperimentData}.

\paragraph{Model instantiation and simulation parameters}
\label{sec:gqlink_helix_params}

For this experimentation, we set the magnetic field density $B$ to 1 tesla and the coefficient $w$ to 0.01. Thus, the particle trajectory has a radius of about $0.38$ mm and completes 417 revolutions in $100$ m (which is the default track length we chose). The maximum step length was set to 20 mm (\stepMax~parameter). We selected QSS2 as GQLink's integration method (we verified that the computational overhead of QSS3 due to its higher order significantly degrades the  performance for this scenario).

\paragraph{Validation}
\label{sec:gqlink_helix_validation}

Geant4 simulations were configured with a relative precision $\epsilon = 10^{-6}$ and setting \deltaChord~and \deltaIntersection~to 0.25 mm and $10^{-5}$ mm, respectively. For this configuration, we then found that setting $\dQRel = 5.1 \times 10^{-5}$ and $\dQMin = 5.1 \times 10^{-5}$ mm in GQLink yields a maximum absolute error in the particle position that is always (i.e., for every cube size) lower than that of Geant4 (for any of the two selected steppers), with an average improvement of about 4x. This error is measured as the Euclidean distance between the simulated and the theoretical position at each given time.

\paragraph{Experimental setup}
\label{sec:gqlink_helix_setup}

We swept 190 equidistant cube sizes in the range\\
$[0.01,1.9]$ mm for side length and, for each size, we characterized the performance of the three methods using the following metrics (cf. Figure \ref{fig:stepping_div}):
\begin{itemize}
    \item The \textsl{end-to-end speedup} against a reference Geant4 stepper, for which we chose \DOPRI as it became recently the default stepper,
    
    \item The \textsl{average \textbf{trajectory calculation} time}, which indicates the average CPU time per step taken by the trajectory calculation and the numerical integration routines. It excludes the intersection-finding algorithms (i.e., it is the time taken by \ComputeStep~subtracting the time taken by \EstimateIntersectionPoint), and
    
    \item The \textsl{average \textbf{boundary crossing detection} time}, which is the average CPU time per volume boundary crossing consumed by \EstimateIntersectionPoint.
\end{itemize}

For each cube size and each metric we ran 20 independent simulations and plotted the average values. The sample standard deviation (vertical bars) remained below 11\% (typically around 5\%). \ref{sec:measurements} provides additional information regarding the time measurements involved in this experimentation.

For the sake of clarity, in what follows we measure \textsl{speedups} preferably in percentage units, according to Equation \ref{eq:speedup_percent}.
\begin{eqnarray}
\label{eq:speedup_percent}
    \textrm{Speedup [times]} &=& \textrm{duration}_\textrm{prev} \, /  \, \textrm{duration}_\textrm{new} \nonumber \\ 
    \textrm{Speedup [\%]} &=& \left(\textrm{Speedup [times]} - 1\right) \times 100
\end{eqnarray}

\paragraph{Performance comparison}
\label{sec:gqlink_helix_perf}

\begin{figure}[hbt]
\centerline{
    \includegraphics[scale=0.5]{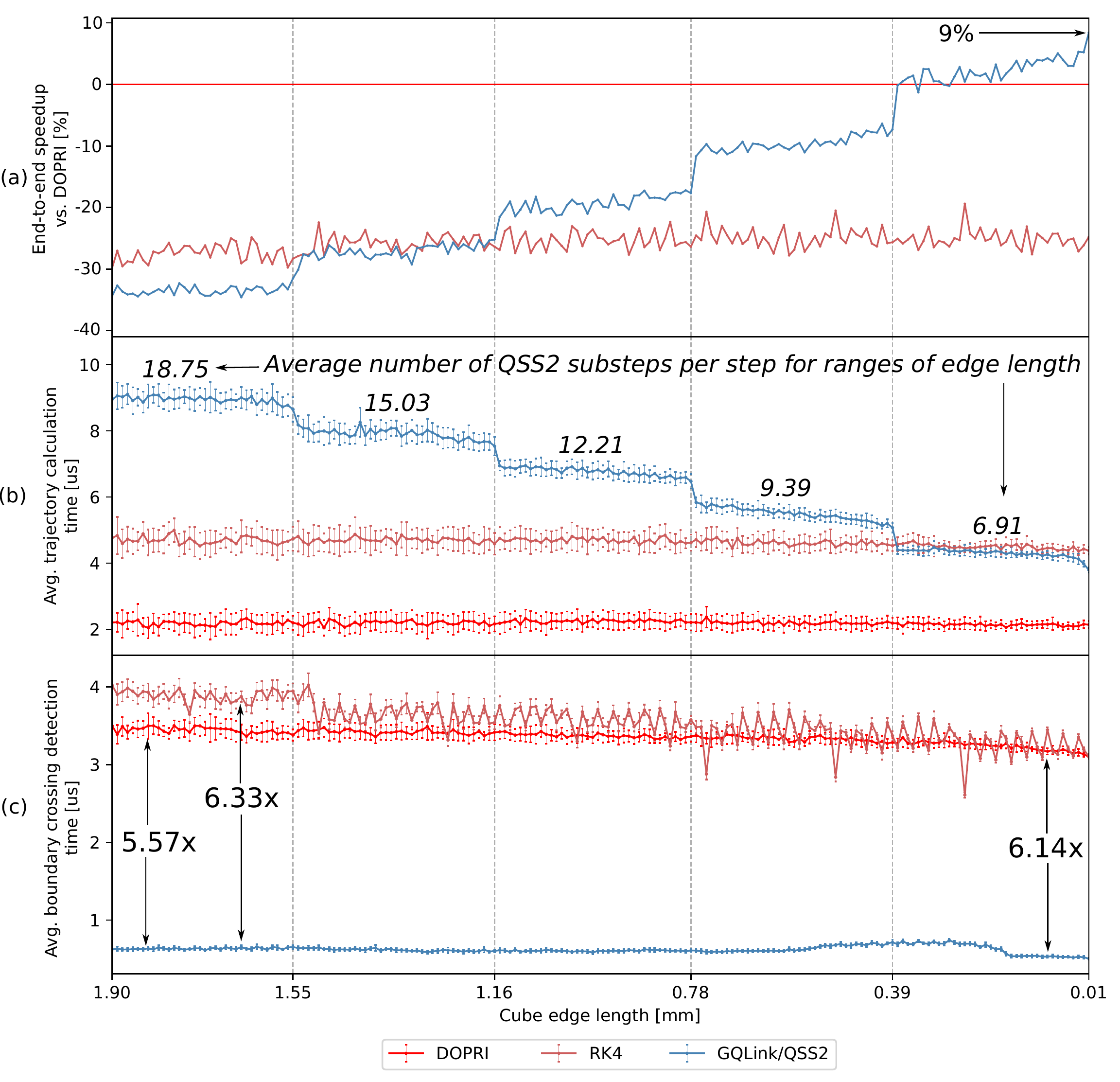}
}
    \caption{Performance comparison between GQLink/QSS2 and Geant4 steppers in case study B2h ($w = 0.01$; $B = 1$ T; $\texttt{stepMax} = 20$ mm; 100 m of track length)}
    \label{fig:GQLink_perf}
\end{figure}

Figure \ref{fig:GQLink_perf}a shows a clear rising trend in GQLink's speedup (against both steppers) as the cube size decreases, reaching a maximum value of about 9\% and 45\% against \DOPRI and \RK, respectively, when the cube edge is 0.01 mm. Inspecting the speedup curve, we can identify five ranges of cube sizes for which GQLink performs increasingly better. This growth pattern can be explained by examining Figure \ref{fig:GQLink_perf}b. GQLink's average trajectory calculation time within each range systematically drops about 20\% each time, and this is a consequence of a decrease in the average number of QSS2 substeps required to complete a single step (annotated on the curve). These numbers are very close to consecutive multiples of 3, which is not surprising since this is the default size of GQLink's substep block. In other words, each range of cube sizes demands less checkpoints until the volume boundary is found. This showcases a very important feature of GQLink: the ability to interrupt prematurely a step without wasting extra CPU resources in calculations of redundant QSS substeps (as we shall see later, our QSS Stepper needs to compute the full step upfront before boundary crossings are handled). On the other hand, we see that both Geant4 steppers achieve a nearly constant average trajectory calculation time: Geant4 requires the steppers to complete a full step before testing for volume boundary crossings (this will be further elaborated in Section \ref{sec:stepper_simulation}).

As for the time spent resolving boundary crossings, we observe in Figure \ref{fig:GQLink_perf}c that GQLink clearly outperforms both steppers, achieving an average improvement over \DOPRI between 6.14x and 5.57x. This is a consequence of the dense output property of QSS methods and the polynomial approximation of trajectories, leveraged in \GQLinkAdvanceConstrained~to enable a very efficient computation of particle movements in the intersection-finding algorithms. 

Every cube size evaluated in this experiment maximizes the ratio of boundary crossings per step (i.e., every step ends at a volume boundary). Yet, it is in the last range of cube sizes where the average trajectory calculation time achieved by GQLink (about 2x higher than \DOPRI) starts to be outweighed by the $\sim$6x improvement in finding intersection points. This ultimately explains the reported maximum end-to-end speedup of about 9\%.

In this range we can also see that GQLink's trajectory calculation time is slightly lower than \RK's. We empirically found that, on average, GQLink can compute one substep in about 13\% of the time taken by \RK to compute a full step. In other words, one GQLink step composed on average of about 7.7 substeps and one \RK step demand approximately the same CPU time in this scenario. This is consistent with the reported average value of 6.91 substeps for that range. As for \DOPRI, we have that one GQLink substep represents a 25\% of a full step, i.e., at most 4 substeps are needed in order to achieve competitive trajectory calculation times. This motivates the need to improve further GQLink's algorithms.

It is also possible to derive an analytical estimation of the end-to-end speedups between any two steppers leveraging information collected during experimentation. In  \ref{sec:formula} we introduce such a formula along with validations for a broad range of scenarios.

%%%%%%%%%%%%%% STEPPER %%%%%%%%%%%%%%%%%%%
\section{Second approach: an embedded QSS stepper}
\label{sec:stepper}

We just showed that QSS and Geant4 can interact gracefully through GQLink. Our next goal is to embed QSS algorithms directly into the Geant4's transportation engine, circumventing the usage of a general purpose layer of interconnection with external toolkits. We accomplish this by designing and implementing a native QSS stepper along with a dedicated QSS integration driver. Our key motivation is to optimize further the particle propagation efficiency by eliminating the overhead introduced by GQLink's abstractions. Thus, our new QSS Stepper sacrifices GQLink's generality in favor of step computation performance. To this end, we stripped and optimized GQLink's core algorithms until achieving a minimalist QSS implementation that can be wired into the Geant4 particle transportation ecosystem.

\subsection{Stepper design}
\label{sec:stepper_design}

\begin{figure}[h]
\centerline{
    \includegraphics[scale=0.5]{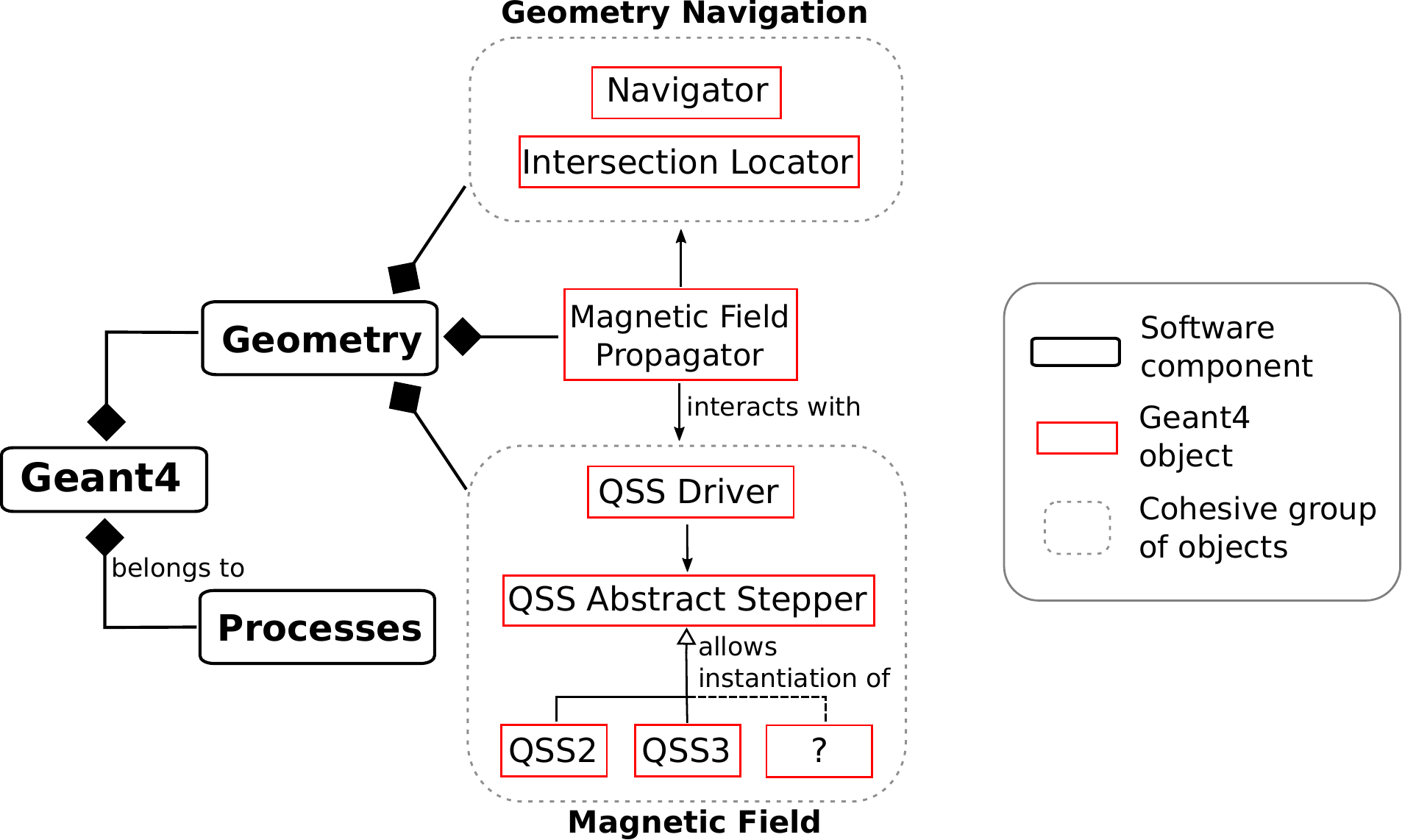}
}
    \caption{High-level structure of Geant4 with the new QSS Abstract Stepper}
    \label{fig:G4_QSS}
\end{figure}

Figure~\ref{fig:G4_QSS} shows the modified Geant4 software component diagram after introducing the new embedded QSS particle propagation capabilities. As opposed to Figure~\ref{fig:GQLink}, we see that the Magnetic Field Propagator no longer interacts with GQLink. The high-level organization is exactly that of Figure \ref{fig:G4}, a consequence of providing a Geant4-compatible implementation of QSS algorithms. The Magnetic Field group now includes a QSS Integration Driver object, a new member of the general Integration Driver class hierarchy, and a QSS Abstract Stepper object (which is in turn part of the Stepper class hierarchy). 

The QSS Driver leverages and reuses most of the behavior of the standard Interpolation Driver, but it also overrides some specific methods, e.g. those that perform adaptive step-size control. Also, it instructs the Stepper to reset the internal QSS integrator state upon starting a new step. This follows the same reasons presented in Section \ref{sec:gqlink_cosim} while discussing GQLink's co-simulation interactions.

Regarding the Stepper, we call it \textsl{abstract} since it offers a single, reusable structure that provides common behavior shared across different \textsl{concrete} algorithms. In spite of this generality, we abide by our performance-driven goals by opting for a generic implementation via C++ templates, which offer negligible runtime performance penalties due to being instantiated at compilation time. As such, the abstract QSS Stepper is created by supplying a QSS Algorithm object as template argument. On runtime, the Stepper just delegates method-dependent work (e.g., right-hand side evaluation of the ODE system) to this object. These delegations correspond to the object interface. Thus, any Algorithm object must implement this interface so that it can interact with the abstract Stepper.

In this work we develop and test embedded implementations of QSS2 and QSS3 methods. We apply them to solve the aforementioned standard equations of motion of charged particles in a magnetic field (Equation \ref{eq:lorentz}). However, other potential members of the QSS family could be smoothly embedded into our framework just by encapsulating the algorithm into an object that implements the delegation interface of the abstract Stepper.

\subsection{Step computation}
\label{sec:stepper_simulation}

\begin{figure}[h]
\centerline{
    \includegraphics[scale=0.35]{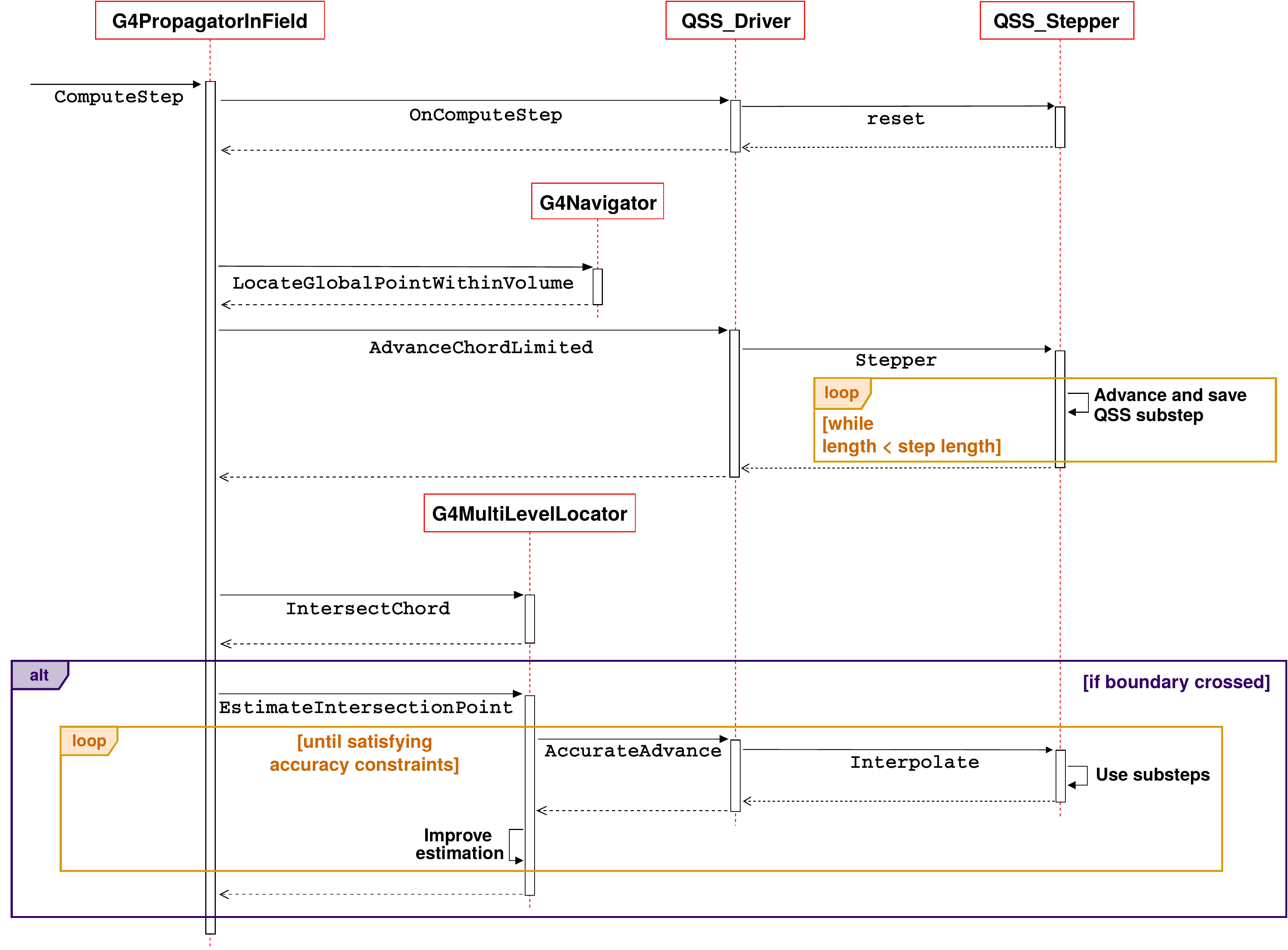}
}
    \caption{Step computation in Geant4 through the QSS Stepper}
    \label{fig:G4_QSS_simulation}
\end{figure}

As shown in Figure~\ref{fig:G4_QSS_simulation}, when Geant4 computes a step, a 
\ComputeStep~call is issued to the Magnetic Field Propagator. It starts by invoking the  \OnComputeStep~method implemented by the Integration Driver in order to perform custom step initialization tasks. There, the QSS Driver issues a \reset~call to prepare the Stepper for the upcoming computation. Basically, the integrator's internal data structures are reinitialized using the particle properties (i.e., charge, mass, velocity and position). This process is an essential part of the particle propagation algorithm since even a single track can exhibit changes in the underlying particle direction or position in-between consecutive steps (as discussed in Section~\ref{sec:gqlink_cosim}).

The particle propagation algorithm continues by locating the starting point of the step in the geometry so that future geometry queries can be properly addressed. Then, the step is actually taken inside the 
\AdvanceChordLimited~routine, implemented by the Integration Driver (the QSS Driver reuses the Interpolation Driver's default implementation, \textit{but skipping the adaptive step size controls} performed once the step is computed). It is interesting to note that the core behavior of the QSS integration routine (\Stepper~method) matches that of GQLink's: each iteration defines a new QSS substep that is stored for later use by the geometry routines. The main difference with GQLink is that the notion of checkpoint no longer exists: geometry boundaries are found once the whole step is taken, and it is the Magnetic Field Propagator's responsibility to do this.

However, the underlying intersection-finding algorithm is essentially equivalent that of GQLink's: it starts with a quick test using a linear segment joining the step endpoints (\IntersectChord), which gives an initial estimation of the intersection point, in case a volume boundary is crossed. This estimation is progressively improved by a more complex algorithm (\EstimateIntersectionPoint) that queries the Integration Driver on each of its iterations (\AccurateAdvance) in order to advance a given length and then test which side of the boundary the particle lies in. The QSS Driver, by means of the Interpolation Driver's custom behavior, issues an \Interpolate~call to the QSS Stepper. This call is handled very efficiently using the QSS polynomials computed previously at each substep. Thus, we see that the \GQLinkAdvanceConstrained~routine discussed in Section \ref{sec:gqlink_cosim} replaces Geant4's \AccurateAdvance.

\subsection{Changes made to Geant4}
\label{sec:stepper_changes}

We engineered the QSS Stepper to demand the least number of changes to Geant4's source code. We extended the \texttt{G4MagIntegrator\-Stepper} and \texttt{G4VIntegration\-Driver} class hierarchies in order to add new classes for our stepper (\texttt{QSS\_Stepper}) and integration driver (\texttt{QSS\_Driver}). We also added a new argument to the \texttt{On\-Compute\-Step} method of the \texttt{G4VIntegrationDriver} class hierarchy in order to reset the internal QSS integrator state before taking each step (as detailed in Section \ref{sec:stepper_simulation}). This argument is a pointer to a \texttt{G4FieldTrack} object.

For convenience purposes, we also implemented a new helper method in the \texttt{G4Mag\-Integrator\-Stepper} class hierarchy. It facilitates the generation of a proper integration driver for the current stepper, and it is invoked during the construction of the Chord Finder.

\subsection{Simulation experiments and discussion}
\label{sec:stepper_results}

We will characterize the performance of the QSS Stepper in an instance of case study B2h (Section~\ref{sec:stepper_results_helix}) and in the context of the case study that models the CMS particle detector (Section~\ref{sec:stepper_results_cms}). The hardware and software platform used throughout the experimentation is described in \ref{sec:setup}. The underlying dataset for this analysis can be found in \cite{ExperimentData}.

\subsubsection{Case study B2h}
\label{sec:stepper_results_helix}

\paragraph{Model instantiation and simulation parameters}
\label{sec:stepper_results_helix_model}

We set the magnetic field density $B$ to $0.01$ tesla and the coefficient $w$ to 0.01, which yields a trajectory radius of about $38$ mm. The particle now completes 4 revolutions in $100$ m of track length. In this case, \stepMax~was set to 0.7 mm, a reference value taken from the average step length computed in realistic CMS simulations.

This parameterization yields particle trajectories that are stepwise ``smooth'', in the sense that -on average- can be accurately approximated by a small number of polynomial QSS substeps (per Geant4 step). This relates directly with the noticeable decrease in the number of revolutions with respect to the experiment covered in Section \ref{sec:gqlink_results}.

\paragraph{Validation}
\label{sec:stepper_results_helix_validation}

As explained in Section \ref{sec:gqlink_helix_validation}, for a given stepper, we measure the absolute error as the Euclidean distance between the simulated and the theoretical position of the particle at each time. In this opportunity, we ensured that the maximum of such errors achieved by the QSS Stepper remains within a 2\% of the one produced by DOPRI (for every cube size). We accomplished this by instantiating $\epsilon = 10^{-5}$, $\deltaChord = 0.25$ mm and $\deltaIntersection = 10^{-5}$ mm in Geant4 and setting $\dQRel = 4.69 \times 10^{-5}$ and $\dQMin = 4.69 \times 10^{-6}$ mm to control the QSS accuracy. In fact, the error in the particle position for the QSS Stepper is at most $1.85\%$ higher than \DOPRI in this setup. GQLink, on the other hand, achieved better error bounds (on average 5x lower than \DOPRI).

\paragraph{Experiment setup}
\label{sec:stepper_results_setup}

Aside from the three performance metrics introduced in Section \ref{sec:gqlink_helix_setup}, in this experiment we also consider the \textsl{average \textbf{particle propagation} time} (cf. Figure \ref{fig:stepping_div}), which represents the average CPU time per step taken by \ComputeStep~(i.e., taking into account trajectory calculation time as well as boundary crossing detection time). 

As discussed in Section \ref{sec:stepper_simulation}, Geant4 finds boundary crossings once the whole step is taken, which makes the steppers behave differently from GQLink. In this sense, this experiment is organized as follows. Cube sizes are grouped into \textsl{cube buckets} where the ratio of boundary crossings remains essentially constant. Each bucket spans cube sizes in-between consecutive multiples of \stepMax. Given a cube with an edge length $l$ such that $k \cdot \stepMax \leq l < (k+1) \cdot \stepMax$, Geant4 demands approximately $k$ full steps inside this cube before detecting the volume boundary and entering the neighboring cube. This yields a $1/(k+1)$ crossing ratio for the corresponding cube bucket.

We swept four consecutive buckets where the crossing ratios increase from $~25\%$ to $~100\%$, sampling 20 equidistant cube sizes within each bucket.

As in the case of the experiment described in Section \ref{sec:gqlink_results}, for each cube size and each performance metric we ran 20 independent simulations and plotted their average values. The accompanying error bars indicate the standard deviation for groups of 20 samples (which was again typically below $11\%$).

\paragraph{Performance comparison}
\label{sec:stepper_results_helix_perf}

\begin{figure}[h!]
\centerline{
    \includegraphics[scale=0.5]{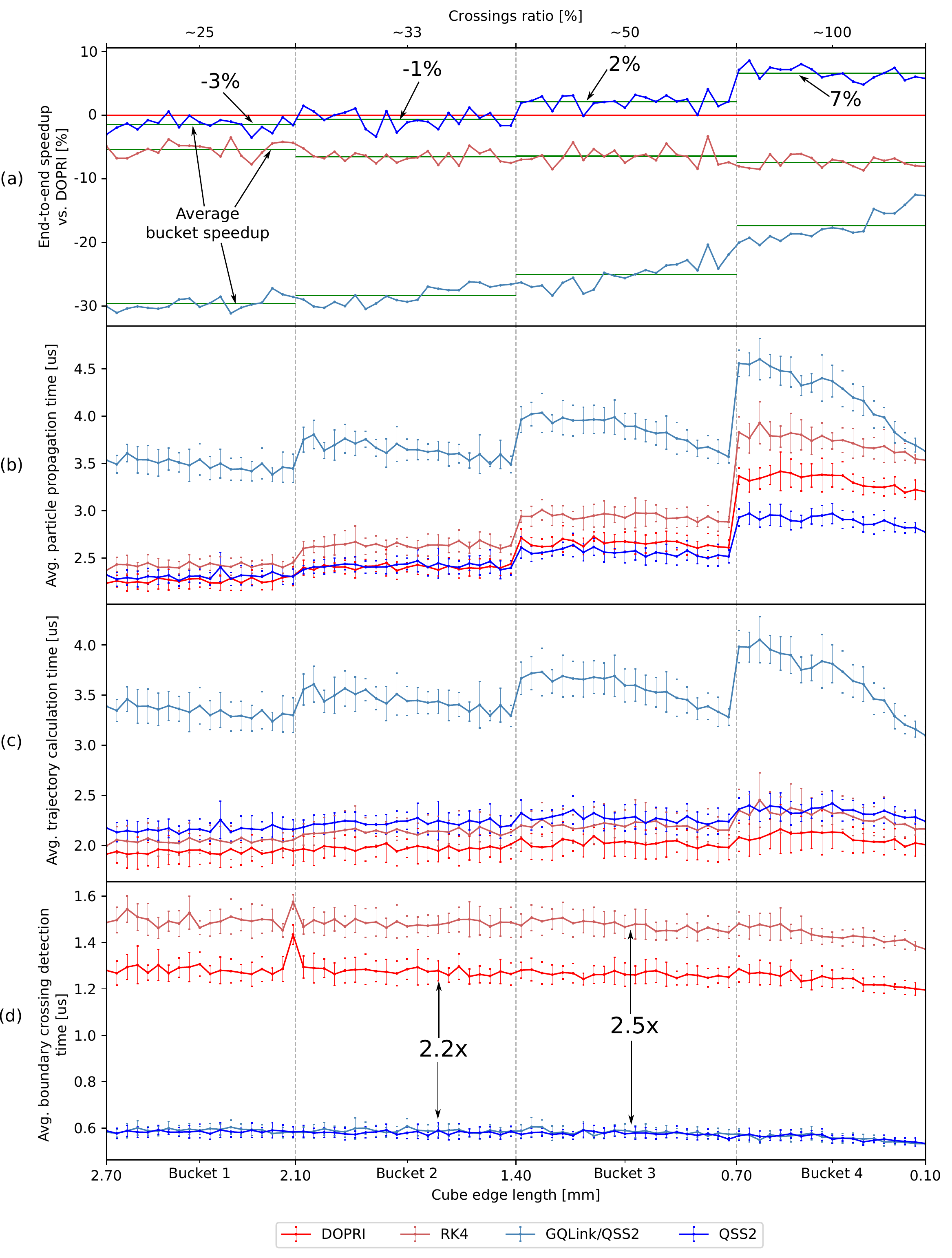}
}
    \caption{Performance comparison between QSS methods and Geant4 steppers in case study B2h ($w = 0.01$; $B = 0.01$ T; $\texttt{stepMax} = 0.7$ mm; 100 m of track length)}
    \label{fig:Stepper_perf_B2h}
\end{figure}

Figure \ref{fig:Stepper_perf_B2h}a shows the end-to-end speedup of each method against Geant4's recent default stepper, \DOPRI. The horizontal green lines represent the average speedup of each method within a given bucket. We can see that QSS2's speedup systematically increases as we move from bucket to bucket, achieving a nearly equivalent performance in bucket 2 (33\% of boundary crossings) and reaching a maximum value of about $7\%$ when the crossings ratio is maximized in bucket 4 (an absolute maximum of $9\%$ is reached within this bucket when the cube edge is approximately of $0.65$ mm). Comparing against \RK, the maximum average bucket speedup is $15\%$ with a absolute maximum of $19\%$. 

As shown in Figure \ref{fig:Stepper_perf_B2h}c, even though the QSS2 trajectory calculation time is consistently $12\%$ above \DOPRI values, the QSS Stepper deals more efficiently with an increasing boundary crossing ratio. In this scenario, both QSS strategies outperform \DOPRI and \RK by a factor of 2.2x and 2.5x, respectively, when computing intersection points (Figure \ref{fig:Stepper_perf_B2h}d). Not surprisingly, the curves for both QSS methods in this Figure are hardly distinguishable to the naked eye: the routines at the heart of the intersection finding algorithm (\GQLinkAdvanceConstrained~in GQLink; \Interpolate~in the stepper) share a nearly equivalent implementation.

Particle propagation information is condensed in Figure \ref{fig:Stepper_perf_B2h}b, which shows the average particle propagation time consumed by each method. For each Geant4 stepper, there are no significant fluctuations of the particle propagation time inside each bucket. This is reasonable to expect: as we already discussed, a given Geant4 stepper will behave very similarly when computing steps inside any cube belonging to the same bucket. However, there is a clear increase in the particle propagation time when jumping from bucket to bucket, which is due to an overall increase in the number of steps needed to cover the track length and complete the simulation.
When reaching the last bucket, QSS2's particle propagation time is $14\%$ lower than \DOPRI's. This is consistent with the reported average end-to-end speedup of $7\%$ inside this bucket, as particle propagation takes roughly the $55\%$ of the whole simulation and $14\% \times 55\% = 7.7\%$.

As for GQLink, we can see a very similar rising trend in the end-to-end speedup as reported in Section \ref{sec:gqlink_helix_perf}, even inside any of the four buckets. However, GQLink's trajectory calculation time in this scenario is considerably slower than \DOPRI's: the time to compute one GQLink substep represents a $40\%$ of the time to compute a complete \DOPRI step, as opposed to the $25\%$ previously discussed in Section \ref{sec:gqlink_helix_perf}. The sustained increase in the crossing ratio is now not enough to compensate this extra difference. On the other hand, we found that the time to compute one substep of the QSS2 stepper represents a $25\%$ of the time to compute one \DOPRI step, which shows a clear improvement in the efficiency of the underlying QSS algorithms. In fact, in this scenario, the Stepper computes a QSS substep $60\%$ faster than GQLink, on average.

\subsubsection{CMS detector application}
\label{sec:stepper_results_cms}

In this Section we provide a complementary performance analysis of the QSS2 Stepper in the context of the case study that models the CMS particle detector (introduced in Section~\ref{sec:cms}).

\paragraph{Simulation parameters}
\label{sec:stepper_results_cms_model}

We used a particle gun shooting a single primary $\pi^-$ particle per event with a kinetic energy of 10 GeV and a random direction within the $\eta$-$\phi$ space, with pseudorapidity $\eta \in [\sfrac{-1}{2}, \sfrac{1}{2}]$ and azimuthal angle $\phi \in [-\pi, \pi]$. Every simulation consisted in 500 independent events.

\paragraph{Validation}
\label{sec:stepper_results_cms_validation}

Since there is no closed-form analytic solution available for CMS simulations, validation in this scenario was achieved through statistical tests to ensure the statistical consistency of QSS simulations against a reference Geant4 stepper (\DOPRI). For this purpose, we used the two-sample Kolmogorov-Smirnov test using a significance level $\alpha = 0.01$ to test the number of steps and tracks produced for all particles tracked in the application (i.e., negative and positive pions, electrons, positrons, gamma photons and a common type for every other particle that might be generated).

\begin{figure}[h!]
\centerline{
    \includegraphics[scale=0.6]{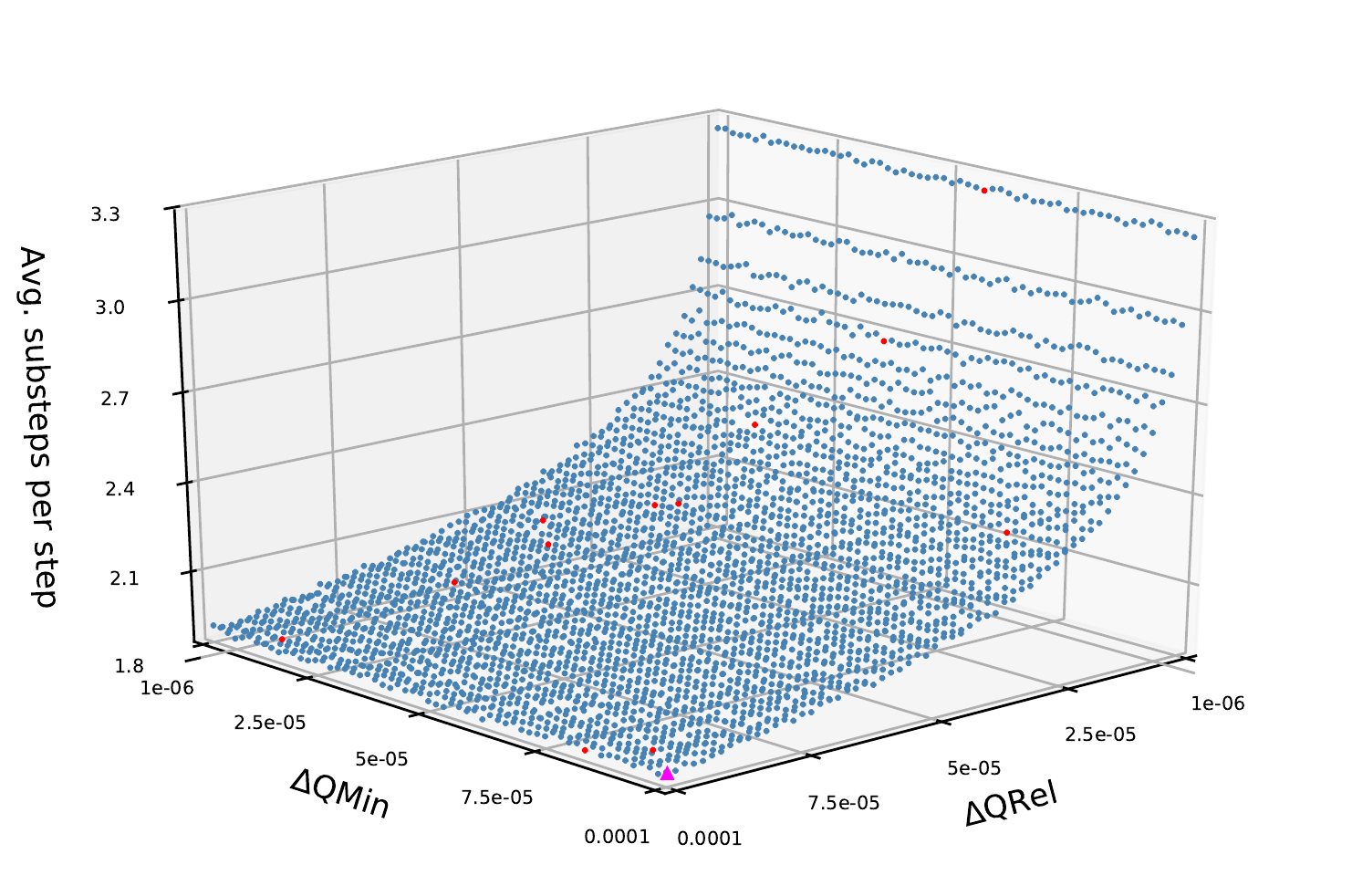}
}
    \caption{Accuracy parameter sweeping for QSS in the CMS setup}
    \label{fig:cms_sweeping}
\end{figure}

We conducted a parameter sweeping for QSS accuracy in order to find a suitable combination of $\dQRel$ and $\dQMin$ that enables statistical consistency while reducing as much as possible the particle propagation times. Thus, we swept 60 equidistant values in the range $[10^{-6}, 10^{-4}]$ for both accuracy parameters (using a fixed RNG seed) and extracted out the average number of QSS substeps per Geant4 step. The results are presented in Figure \ref{fig:cms_sweeping} (red dots indicate cases of failure in the statistical validation). We used \DOPRI with the same set of Geant4 accuracy parameters as described in Section \ref{sec:stepper_results_helix_validation}. The surface shows, as expected, that the number of substeps increases as the requested relative precision is more stringent, while there are no  significant changes when varying $\dQMin$ (for a fixed $\dQRel$). The minimum number of substeps (about 1.9) is achieved at $\dQRel = \dQMin = 9.8322 \times 10^{-5}$ (marked with a pink triangle). We selected this accuracy to carry out the performance comparison, as it passed the statistical tests.

\paragraph{Experiment setup}
\label{sec:stepper_results_cms_setup}

We selected 100 random seeds to initialize Geant4's RNG and, for each one of them, we ran 10 independent simulations to measure the end-to-end simulation time achieved by each stepper and other 10 independent simulations to measure the average particle propagation times per step. These metrics are summarized with average values for groups of 10 samples.

\paragraph{Performance comparison}
\label{sec:stepper_results_cms_perf}

\begin{figure}[hbt]
\centering
  \subfloat[QSS2 vs. \DOPRI\label{fig:cms_dopri}]
    {
        \includegraphics[scale=0.57,valign=t]{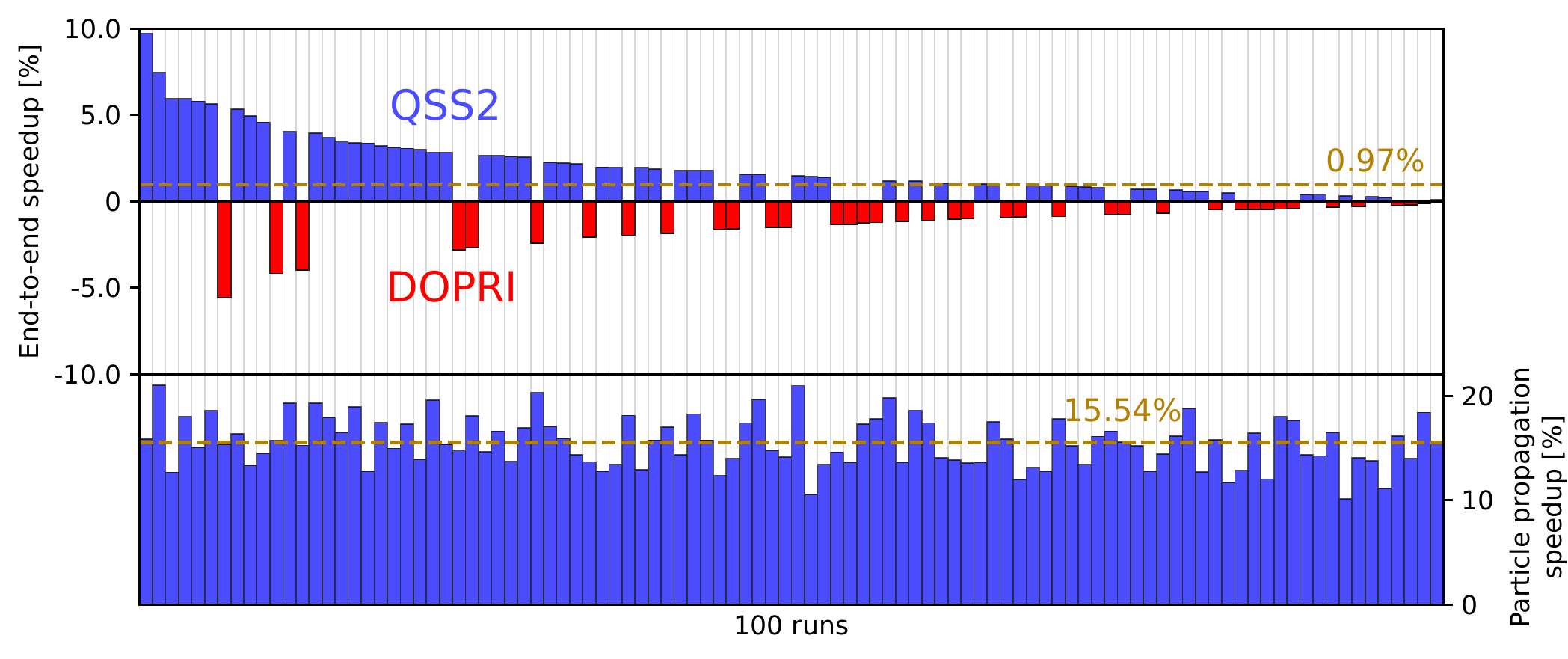}
    }
  \hfill
  \subfloat[QSS2 vs. \RK\label{fig:cms_rk4}]
    {
        \includegraphics[scale=0.57,valign=t]{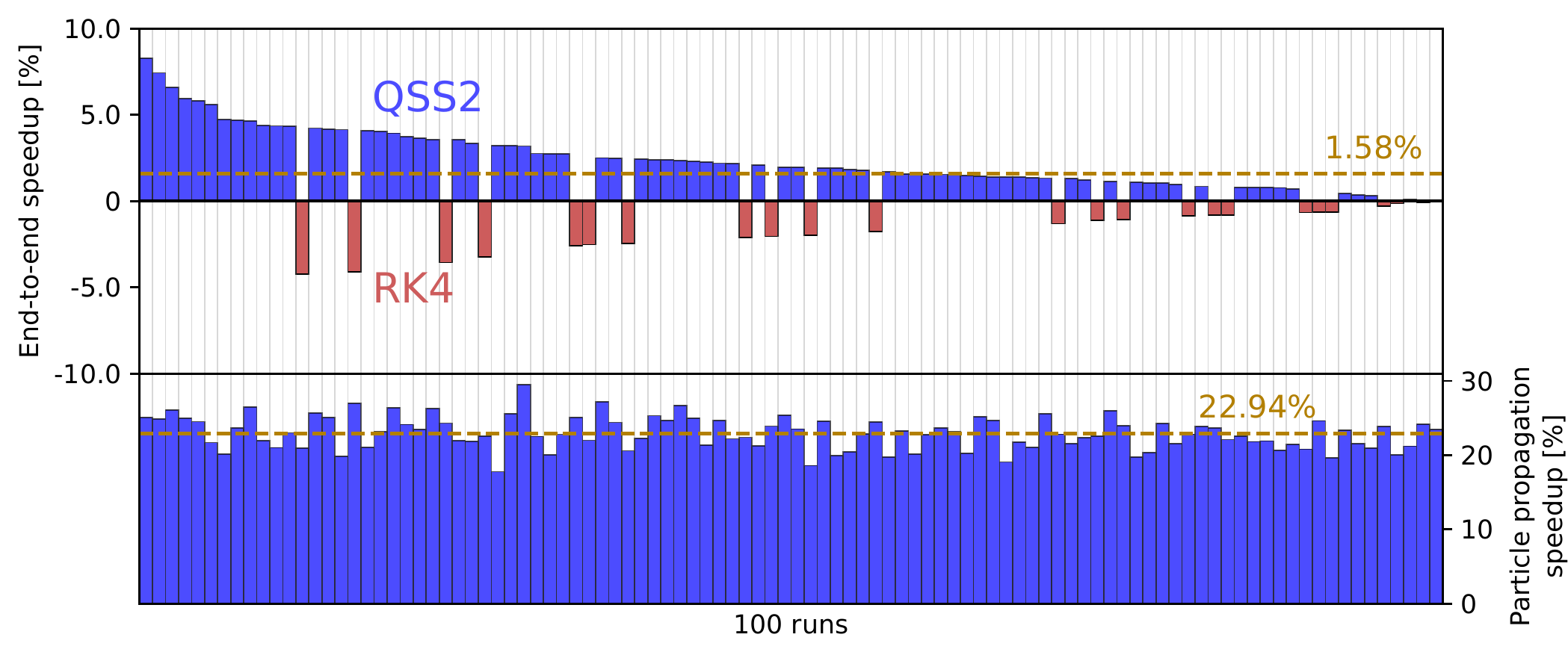}
    }

\caption{Performance comparison for the CMS case study (500 single $\pi^-$ events for each run)}
\end{figure}

In Figures \ref{fig:cms_dopri} and \ref{fig:cms_rk4} we present the performance comparison between the QSS2 Stepper and \DOPRI and \RK, respectively, for 100 simulation runs. The upper graph shows the relative end-to-end speedups between both steppers, whereas the lower complements this information providing the corresponding particle propagation speedup. Simulations are sorted with a decreasing absolute value of the end-to-end speedup.

We found that 62 simulations favored QSS2 against \DOPRI (positive blue bars in the upper graph of Figure \ref{fig:cms_dopri}). This number increases to 77 when comparing against \RK. Notably, on the other hand, the QSS2 Stepper offered significant performance gains regarding particle propagation times, outperforming both \DOPRI and \RK in every single simulation. The dashed brown lines in the lower graphs show the average particle propagation speedups: about 15\% against \DOPRI and roughly 23\% against \RK. 

In spite of this, the end-to-end speedups occasionally favor the standard Geant4 steppers. This can be explained by a proportionally small CPU demand for particle propagation procedures in CMS compared to CPU demand for physics evaluation: particle propagation takes about 7.5\% of the end-to-end simulation time for \DOPRI and nearly 8\% for \RK. In both situations, the physics processes are the most CPU-intensive software components.

Another salient feature of the CMS scenario is its small ratio of boundary crossings, which is typically between 8\% and 9\% of the total number of steps. Although this is not a property that can be leveraged by QSS methods, the favorable speedups observed can be  explained by two facts: first, the underlying trajectories tend to be stepwise smooth, demanding less than 2 QSS substeps per step, on average, as discussed in Section \ref{sec:stepper_results_cms_validation}. Second, we found that the time to compute one QSS2 substep represents on average a 43\% of a complete \DOPRI step, and a 40\% of an \RK step. Thus, particle propagation speedups against each stepper should be expected for an average number of QSS substeps below 2.32 and 2.5, respectively. With an average particle propagation speedup of 15\% against \DOPRI, and considering that particle propagation takes 7.5\% of the end-to-end simulation time, we would expect an average end-to-end speedup of $15\% \times 7.5\% \approx 1.1\%$. This figure is close to the reported average end-to-end speedup of 0.97\% (dashed brown line in the upper graph of Figure \ref{fig:cms_dopri}). The same reasoning yields an expected end-to-end speedup of 1.8\% against \RK. In this case, the empirical end-to-end average speedup was 1.58\% (dashed brown line in the upper graph of Figure \ref{fig:cms_rk4}).

In order to put these figures into perspective, we claim that a strict upper bound for the expected end-to-end speedup in the CMS application should be between 7.5\% and 8\%, as these are the average ratios of CPU resources taken by the two Geant4 steppers we analyzed. Bearing this in mind, the reported end-to-end speedups of 0.97\% and 1.58\% represent a 13\% and a 20\% of those upper bounds. In addition, as already discussed in Section \ref{sec:intro}, a 1\% reduction in CMS simulation times can yield very important savings.

Finally, we found that the QSS2 Stepper computes a QSS substep about 42\% faster than GQLink, on average. In this case, the time to compute a single substep represents a 37\% of the time to compute a full GQLink step.

As a final note, this performance comparison is solely based on $\pi^-$ events. Other types of particles could give different results, as e.g. muons will traverse more volumes and cross more boundaries. Thus, we stress the need of developing similar studies using simulated collision events of different types, which are part of upcoming research efforts.

%%%%%%%%%%%%% CONCLUSIONS %%%%%%%%%%%%%%%
\section{Concluding remarks}
\label{sec:conclusions}

We studied the applicability of state quantization-based numerical solvers in the simulation of moving particles and their interaction with traversed matter. In particular, we focused in the domain of High Energy Physics (HEP) experiments that deals with the behavior of subatomic particles. These problems involve solving continuous equations of motion to track particle trajectories within complex detector geometries made up of 3D volumes. This suggests the appropriateness of numerical methods that deal efficiently with discontinuity handling in continuous models, a situation represented by the crossing of boundaries between adjacent volumes by a travelling particle. We adopted the QSS family of numerical solvers that are designed in the realm of a discrete-event simulation approach, making discontinuity detection and handling a cheap procedure as compared to classical discrete-time oriented solvers (e.g. the Runge-Kutta family of methods).

We developed several tools and models to apply QSS for HEP experiments and analyzed their performance in varied case studies, from abstract simplified models to real-world complex particle detectors. 

The first simple case is a basic HEP setup consisting of a single positron describing a circular 2D trajectory under the action of an uniform magnetic field (B2c). Performance and accuracy comparisons were made between Geant4 (the most widely used simulation toolkit in modern HEP experiments) and QSS Solver (the state-of-the-art toolikt for QSS methods). We confirmed that QSS can substantially profit from scenarios where the number of boundary crossings is high (modeled in this case as parallel planes placed across the trajectory of the particle): QSS Solver performed up to 6 times faster than Geant4 with better error bounds. Yet, it is very difficult to apply directly QSS Solver to simulate complex realistic HEP problems, as it is a general purpose simulation toolkit (e.g. simulating complex stochastic physics processes is out of its scope).

Therefore, we developed a co-simulation scheme, GQLink, to combine the two simulation toolkits while preserving the independence of their core engines. GQLink takes the responsibility of particle tracking via ODE solving, while Geant4 remains in charge of driving the simulation and solving stochastic particle-matter physics interactions. GQLink proved effective as an abstract mechanism to connect Geant4 to arbitrary external simulation toolkits. 

Finally, we engineered a minimalist version of QSS methods natively embedded into the Geant4 transportation engine: the QSS Stepper. It provides a common integration scheme suitable for different kinds of QSS methods, with negligible performance penalties. So far, we developed and tested two implementations for QSS2 and QSS3. The know-how and lessons learned from the deployment of GQLink proved to be essential to facilitate and boost this development process, which can be leveraged for the design of future solvers.

We analyzed both strategies, GQLink and QSS Stepper, in the context of two complementary case studies: a synthetic representative setup where a positron describes a 3D helical trajectory within a lattice of cubes (B2h), and a real-world scenario based on the Compact Muon Solenoid (CMS) particle detector. We established performance  comparisons against the two most relevant Geant4 steppers, the \RK and \DOPRI discrete-time methods, revealing that the end-to-end performance is scenario-dependent, as each approach has its own, distinctive strengths. 

In scenarios with heavy volume crossing activity, stepwise smooth trajectories (i.e., each step being composed on average of a small number of QSS substeps) are typically well suited for the QSS Stepper.

We found an instantiation of the B2h case study where QSS2 outperforms the standard Geant4 steppers when the ratio of boundary crossings is greater than 33\%, achieving speedups of up to 7\% and 15\% against \DOPRI and \RK when this ratio is maximized. 

On the other hand, non-stepwise smooth trajectories can be best capitalized by GQLink, as its particle propagation capabilities allow for prematurely interrupting the step computation as soon as a boundary crossing is found. For example, in an alternative parameterization of the B2h case study featuring a more intense magnetic field, GQLink systematically improves its end-to-end simulation performance as cube sizes decrease, achieving speedups of up to 9\% and 45\% against \DOPRI and \RK, respectively.

In scenarios that feature a very small proportion of volume crossings, QSS can still offer performance gains, as confirmed by the CMS case study. Although the ratio of volume crossings to integration  steps is between 8\% and 9\%, we found that trajectories in this setup tend to be stepwise smooth: their steps are generally composed of a number of QSS substeps that is small enough to enable the QSS2 Stepper to outperform both \DOPRI and \RK, achieving average end-to-end speedups of $\sim$1\% and $\sim$1.5\%, respectively. Since the particle propagation routines in this scenario consume between $7.5\%$ and $8\%$  of the total CPU time, these performance improvements stand for $13-20\%$ of the upper bound theoretically achievable.

We are interested in testing our strategies in the context of other realistic HEP setups. To this end, we are currently developing a standalone Geant4 application that models the ATLAS detector at CERN \cite{Aad2008ATLAS}. It would also be important to extend the experiments of the CMS case study considering other types of particles beyond $\pi^-$. This extended study may provide a more solid and complete characterization of QSS for use in production CMS simulations.

We are also working on enhancing QSS Solver to perform particle tracking in a broader range of application domains not limited to HEP, implementing efficient geometry control as particles move throughout 3D spaces composed of faceted polyhedrons. These applications include plasma simulation and transport of pollutant particles affecting air quality in urban zones.

\section*{Acknowledgments}

The authors want to greatly thank Dr. Soon Yung Jun, Dr. Krzysztof Genser, Dr. Daniel Elvira (Fermi National Accelerator Laboratory, Chicago, USA) and Prof. Ernesto Kofman (Universidad Nacional de Rosario and CONICET, Argentina) for their valuable comments and insights during the development of this manuscript. This paper does not fall within the publication policy of the Geant4 Collaboration.

This work was partially supported by the National Agency for Science and Technology (ANPCYT, grant PICT-2015-3509) and the University of Buenos Aires (UBACYT PhD Fellowship Program).

%%%%%%%%%%%% APPENDIX %%%%%%%%%%%
\appendix

\section{Hardware and software platform}
\label{sec:setup}

All simulations were run on the computer cluster TUPAC \cite{TUPAC}, where each CPU node has 4 x AMD Opteron 6276 (hexadeca-core) processors. The operating system in use is Red Hat Enterprise Linux ComputeNode release 6.7 (\texttt{2.6.32-573.el6.x86\_64} kernel).

We used Geant4 version 10.5 \cite{G4Release1005}, the latest official release as of the writing of this article, compiled with \texttt{gcc} 5.4.0 in release mode (i.e., with optimization flags turned on). GQLink and the QSS Stepper were based upon the QSS Solver engine from version 3.0. Each simulation was single-threaded.

\subsection{Time measurements}
\label{sec:measurements}

Three independent Geant4 release-mode builds were used to extract assorted statistics, including the different performance metrics reported in the article. Each build was produced using a custom \texttt{cmake} option to enable time measurement and \textsl{bookkeeping} code upon compilation:
\begin{itemize}
    \item The \textsl{NoStats} build is the regular, unmodified build we used to extract the \textsl{end-to-end simulation times} (which are in turn used to compute the \textsl{end-to-end speedups}).
    
    \item The \textsl{ComputeStepStats} build enables code to measure the time spent inside the \texttt{Compute\-Step} method of the Magnetic Field Propagator. This build is used to calculate the average \textsl{particle propagation times}.
    
    \item The \textsl{BoundaryCrossingStats} build enables code to measure the time spent inside \texttt{Estimate\-Intersection\-Point}. The average \textsl{boundary crossing detection times} are calculated using this build.
\end{itemize}

As explained in Section \ref{sec:gqlink_helix_setup}, the \textsl{trajectory calculation time} is computed as the difference between the \textsl{particle propagation time} and the \textsl{boundary crossing detection time}.

Each Geant4 simulation was run once per each of these three builds. In turn, as detailed in Sections \ref{sec:gqlink_helix_setup}, \ref{sec:stepper_results_setup} and \ref{sec:stepper_results_cms_setup}, the experiments consisted in $N$ independent repetitions of a single simulation ($N = 20$ for case study B2h and $N = 10$ for the CMS application).

\section{Analytic estimation of speedups}
\label{sec:formula}

Given two integration methods $m_1$ and $m_2$, we wish to calculate the expected end-to-end speedup of $m_1$ (conceived as the new, alternative method) over $m_2$ (conceived as the reference baseline method) in the context of a certain scenario $\Sigma$. By \textsl{method} we understand any Geant4 stepper or also GQLink equipped with any of its underlying numerical solvers, whereas a \textsl{scenario} is a simulation setup such as the CMS application. 

Our proposed speedup estimation formula is defined in terms of four components, all of them under scenario $\Sigma$:
\begin{itemize}
    \item The \textsl{trajectory calculation speedup} $\SteppingSpeedup = \SteppingSpeedup_\Sigma(m_1,m_2)$ of $m_1$ over $m_2$, defined as the quotient between $m_2$ and $m_1$ trajectory calculation times,
    
    \item The \textsl{boundary crossing detection speedup} $\CrossingSpeedup = \CrossingSpeedup_\Sigma(m_1,m_2)$ of $m_1$ over $m_2$, defined as the quotient between $m_2$ and $m_1$ boundary crossing detection times,
    
    \item The $m_2$ \textsl{particle propagation ratio} $c = c_\Sigma(m_2)$ which is the proportion of CPU time consumed by the particle propagation routines of $m_2$, and    
    
    \item The $m_2$ \textsl{boundary crossing detection ratio} $s = s_\Sigma(m_2)$, which is the ratio between the CPU time spent in $m_2$ boundary crossing detection routines and the CPU time of all $m_2$ particle propagation routines.
\end{itemize}

The formula is presented in Equation \ref{eq:speedup}. Intuitively, the numerator represents a complete $m_2$ simulation as one unit, whereas the denominator estimates the proportion of that unit that represents a complete $m_1$ simulation. In turn, we decompose the $m_2$ simulation unit into a trajectory calculation unit and a boundary crossing detection unit and we compute for each one of them the corresponding proportions that arise when switching to $m_1$. If for example $m_1$ has a boundary crossing detection speedup over $m_2$ of 4x, for a given intersection point we can think that the former only consumes 25\% of the time consumed by the latter. Thus, the remaining 75\% is spared and is accounted by multiplying it with the proportion of the overall CPU time spent solving boundary crossings ($c \cdot s$).

\begin{equation}
    \Speedup_\Sigma(m_1, m_2) = \frac{1}{1 - \left(1 - \sfrac{1}{\SteppingSpeedup}\right) \cdot \left((1 - c) \cdot s\right) - \left(1 - \sfrac{1}{\CrossingSpeedup}\right) \cdot c \cdot s}
    \label{eq:speedup}
\end{equation}

%\subsection{Validation}
%\label{sec:formula_results}

Table \ref{tab:speedups} shows a comparison between end-to-end speedups found empirically and analytically via Equation \ref{eq:speedup} for three different scenarios: two instances of the B2h case study (B2h-a and B2h-b) and the CMS application as studied in Section \ref{sec:stepper_results_cms}. B2h-a corresponds to the fourth cube bucket introduced in Section \ref{sec:stepper_results_helix_perf} (i.e., 100\% of boundary crossings), whereas B2h-b focuses on the third cube bucket setting $B = 1$ tesla. For each scenario, the four parameters required by the formula were computed taking as input the particle propagation information found empirically.

We can see that the speedup formula yields practical orientative  predictions, with a worst case error of 25\%, but typically lower. In fact, for a broader range of scenarios and different combinations of methods (more than 100 cases), we found that the average error is 12\% with a standard deviation of 7\%.

% helix 1:  b=0.01, vx=0.01, bucket 2
% helix 2: b=1, vx=0.01, bucket 1
\begin{table}[ht]
\begin{center}
\caption{Comparison of empirical and analytical end-to-end speedups\label{tab:speedups}}
%\hspace{-2.5cm}
\begin{small}
%\begin{tabular}{| c | ccc | ccc | ccc |}
\begin{tabular}{| c | ccc | ccc |}
    \hline
    \multicolumn{1}{|c|}{\textbf{Scenario}} &
    \multicolumn{3}{c|}{\textbf{Speedup QSS2 vs. \DOPRI}} &
    %\multicolumn{3}{c|}{\textbf{QSS2 vs. RK4}} &
    \multicolumn{3}{c|}{\textbf{Speedup \DOPRI vs. \RK}}\\
        
    & Empirical & Expected & Error &
    %Empirical & Expected & Error &
    Empirical & Expected & Error \\
    \hline
    
    %B2h-a & 2.14\% & 2.05\% & 4.09\% & 9.16\% & 8.01\% & 12.53\% & 6.88\% & 5.86\% & 14.79\%\\
    %B2h-b & -16.81\% & -16.46\% & 2.13\% & -10.83\% & -11.98\% & 10.62\% & 7.28\% & 5.94\% & 18.31\%\\
    %CMS & 0.97\% & 1.02\% & 6.14\% & 1.58\% & 1.51\% & 4.34\% & 0.64\% & 0.48\% & 25\%\\
    
    B2h-a & 7.02\% & 7.94\% & 13.1\% & 8.01\% & 7.23\% & 9.87\%\\
    B2h-b & -12.38\% & -11.92\% & 3.74\% & 7.28\% & 5.94\% & 18.31\%\\
    CMS & 0.97\% & 1.02\% & 6.14\% & 0.64\% & 0.48\% & 25\%\\
    \hline
\end{tabular}
\end{small}
\end{center}
\end{table}

Consider as an example the CMS application. In Section \ref{sec:stepper_results_cms_perf} we discussed that \DOPRI-enabled simulations spend roughly a 7.5\% of the time in particle propagation routines, i.e., $s = 0.075$. We also found empirically that the boundary crossing detection ratio $c = 0.05$, i.e., a 5\% of the particle propagation time is devoted to detect intersection points. Also, the trajectory calculation speedup $\sigma$ of the QSS2 Stepper over \DOPRI was found to be $1.13$x, whereas the boundary crossing detection speedup $\gamma = 2$x. Thus, we have that

$$\Speedup_{\textrm{CMS}}(\textrm{QSS2}, \textrm{DOPRI}) \approx 1.01017 = 1.017\%$$

Equation \ref{eq:speedup} can have several interesting applications. For example, it can be used to drive the design of future steppers, setting target speedup values and testing whether the expected end-to-end speedup against a reference stepper in a case study of interest is within an acceptable range.

\end{document}